\documentclass{aa}  

\usepackage{graphicx}
\usepackage{txfonts}
\usepackage{xcolor}
\usepackage{appendix}
\usepackage{placeins}
\usepackage{longtable}
\usepackage{soul}
\usepackage[mathscr]{euscript}
\usepackage[colorlinks=true,citecolor=blue]{hyperref}
\graphicspath{{./}{plots/}}

\begin{document} 

   \title{Multi-campaign Asteroseismic Analysis of eight Solar-like pulsating stars observed by the K2 mission}
\titlerunning{Multi-campaign Asteroseismic Analysis of eight K2 Solar-like stars}

   \author{L. Gonz\'alez-Cuesta
          \inst{1,2}
          \and
          S. Mathur\inst{1,2}
          \and
          R. A. Garc\'ia\inst{3}
          \and
          F. P\'erez Hern\'andez\inst{1,2}
          \and
          V. Delsanti\inst{1,2,3,4}
          \and
          S.N. Breton\inst{3}
          \and
          C. Hedges\inst{5,6}
          \and
          A. Jim\'enez\inst{1,2}
          \and
          A. Della Gaspera\inst{3,4}
          \and
          M. El-Issami\inst{3,4}
          \and
          V. Fox\inst{3,4}
          \and
          D. Godoy-Rivera\inst{1,2}
          \and
          S. Pitot\inst{3,4}
          \and
          N. Proust\inst{3,4}
          }

   \institute{Instituto de Astrof\'isica de Canarias (IAC), V\'ia L\'actea S/N, La Laguna 38200, Tenerife, Spain\\
              \email{luciagc@iac.es}
         \and
             Departamento de Astrof\'isica, Universidad de la Laguna, La Laguna 38200, Tenerife, Spain 
        \and
            Universit\'e Paris-Saclay, Universit\'e Paris Cité, CEA, CNRS, AIM, 91191, Gif-sur-Yvette, France
        \and
           Ecole Centrale-Supelec, Universit\'e Paris-Saclay, 91190 Gif-sur-Yvette, France
        \and
            Bay Area Environmental Research Institute, P.O. Box 25, Moffett Field, CA 94035, USA
        \and
            NASA Ames Research Center, Moffett Field, CA 94035, USA
             }



 
  \abstract
    {The NASA K2 mission that succeeded the nominal {\it Kepler} mission observed several hundreds of thousands of stars during its operations. While most of the stars were observed in single campaigns of $\sim$\,80 days, some of them were targeted for more than one campaign. We perform an asteroseismic study of a sample of eight solar-like stars observed during K2 Campaigns 6 and 17, allowing us to have up to 160\,days of data. With these two observing campaigns we determine not only the stellar parameters but also study the rotation and magnetic activity of these stars. We first extract the light curves for the two campaigns using two different pipelines, \texttt{EVEREST} and \texttt{Lightkurve}. The seismic analysis is done on the combined light curve of C6 and C17 where the gap between them was removed and the two campaigns were stitched together.
    We determine the global seismic parameters of the solar-like oscillations using two different methods: the A2Z pipeline and the Bayesian \texttt{apollinaire} code. With the latter, we also perform the peak-bagging of the modes to characterize their individual frequencies. By combining the frequencies with the {\it Gaia} DR2 effective temperature and luminosity, and metallicity for five of the targets, we determine the fundamental parameters of the targets using the IACgrids based on the MESA (Modules for Experiments in Stellar Astrophysics) code. 
    We find that four of the stars are on the main sequence, 
    two stars are about to leave it, and two stars are more evolved (a subgiant and an early red giant). While the masses and radii of our targets probe a similar parameter space compared to the {\it Kepler} solar-like stars with detailed modeling, we find that for a given mass our more evolved stars seem to be older compared to previous seismic stellar ensembles. 
    We calculate the stellar parameters using two different grids of models, incorporating and excluding the treatment of diffusion, and find that the results agree generally within the uncertainties, except for the ages. The ages obtained using the non-diffusion models are older with differences greater than 10\% for most stars. The seismic radii and the {\it Gaia} DR2 radii present an average difference of 4\% with a dispersion of 5\%. Although the agreement is quite good, the seismic radii are slightly underestimated compared to {\it Gaia} DR2 for our stars, the disagreement being greater for the more evolved ones. The rotation analysis provides two candidates for potential rotation periods but longer observations are required to confirm them.}


   \maketitle
%

\section{Introduction}



Thanks to data collected by missions such as CoRoT \citep[Convection, Rotation, and Transits,][]{2006cosp...36.3749B} and {\it Kepler}/K2 \citep{2010Sci...327..977B,2014PASP..126..398H}, asteroseismology has demonstrated that it is a powerful tool to determine more precise stellar parameters compared to classical methods and provide information on stellar interiors \citep[e.g.][]{2021RvMP...93a5001A}.


For stars like the Sun, with an internal radiative zone and a convective envelope, solar-like oscillations are generated by the turbulent motions in the outer layers of the star, yielding stochastically-excited modes. The excellent precision of the {\it Kepler} observations led to many asteroseismic studies of stars with gravito-acoustic oscillations. Indeed, solar-like oscillations have been detected and characterized in hundreds of stars on the main sequence and on the subgiant branch \citep[e.g.][]{2009A&A...506...41G,2017ApJS..233...23S,2019LRSP...16....4G,2021FrASS...7..102J} as well as in tens of thousands of red giants \citep[e.g.][]{2018ApJS..236...42Y}. In addition several tens of planet-host stars have been characterized with asteroseismology, enabling a more precise measurement of the planets radii and ages \citep[e.g.][]{2013ApJ...767..127H,2015ApJ...799..170C,2015MNRAS.452.2127S}.

The NASA {\it Kepler} Mission ended in May 2013 because of the failure on the second of the four reaction wheels of the satellite. This problem made it impossible to continue with the nominal mission and the situation forced the team to design a new observation strategy, giving rise to the K2 mission \citep{2014PASP..126..398H}. This mission resulted in 20 observation campaigns of around 80 days along the ecliptic plane in different regions of the Galaxy. In the early campaigns of the K2 mission, solar-like oscillations were detected in 36 solar-like stars including 3 planet-host stars \citep{2015PASP..127.1038C,2016PASP..128l4204L,2018MNRAS.478.4866V,2019AJ....158..248L,2021MNRAS.tmp.3152S} as well as in several tens of thousands of red giants \citep{2017ApJ...835...83S,2020ApJS..251...23Z,2022ApJ...926..191Z}. 




There are some overlapping campaigns\footnote{\url{https://archive.stsci.edu/missions-and-data/k2/campaign-fields}}, which enable us to double the observation time for several targets. This is the case for campaigns C6 and C17, yielding a sample of twelve solar-like stars that have been observed in both campaigns in short cadence. Among them, solar-type oscillations have been detected with a signal-to-noise ratio high enough for asteroseismic analysis for eight stars\footnote{These stars are also part of the analysis of a larger set of solar-like oscillators from K2 C6-19 by Lund et al. (in prep.).} and with a {\it Gaia} Renormalized Unit Weight Error (RUWE) value below 1.25 to avoid possible binaries. These are the stars we analyze in this study with asteroseismology. We note that five of our targets (EPIC~212478598, EPIC~212485100, EPIC~212487676, EPIC~212516207 and EPIC~212683142) have also been seismically studied by \citet{2021ApJ...922...18O}. However they only analyzed one campaign (C6). Our work uses time series twice as long, hence with a higher resolution compared to the previous analysis and we show in Section~\ref{sec:PB} that we retrieve more modes at higher and lower frequency as well as higher degree modes. In addition, we present a new asteroseismic analysis for another 3 solar-like stars (EPIC~212509747, EPIC~212617037 and EPIC~212772187).


Note that two other stars show solar-like oscillations in both campaigns. EPIC~212708252 is a solar analog with a RUWE value of 1.274, so slightly above our cut. That star is part of another paper devoted to solar analogs (Garc\'ia et al. in prep.). EPIC~212709737 is a hot F dwarf with a {\it Gaia} effective temperature around 6,500\,K and as expected for such a hot star \citep{2012A&A...537A.134A}, the modes are wide making their characterization more complicated. Besides, with a RUWE value of 2.469 it is more likely to be in a binary system. This star is currently under spectroscopic follow-up requiring a longer timeline for a full  analysis of the system, which is out of the scope of this paper.

The layout of the paper is as follows, in Section 2, we describe the data that we used and the procedure to calibrate the light curves for our asteroseismic studies. In Section 3, we describe the atmospheric parameters of our sample of stars that will be used for the stellar modeling. In Section 4, we explain the procedure followed to characterize the modes and to model the stars. Section 5 discusses the results from the stellar modeling as well as the analysis of the rotation and magnetic activity of the stars. Finally, in Section 6, we provide the conclusions of this work.

\section{Photometric Observations}\label{sec:2}
We select eight stars observed in short cadence mode \citep[SC, $dt\sim$\,1 minute, see for more details][]{2010ApJ...713L.160G} observed in both campaigns, C6 and C17, for which signatures of a p-mode hump were found. The full list of stars, which we will refer to with letters from A to H, including their general properties is shown in Table~\ref{table:1}.


Compared to the nominal {\it Kepler} mission, the K2 observations suffer additional systematics due to the scheme adopted to stabilize the satellite with only two working reaction wheels. Increased spacecraft roll motion around the boresight caused a saw-tooth shaped systematic in mission light curves on time-scales of approximately 6 hours. In order to produce light curves without those systematics and properly calibrate the K2 data, several pipelines were developed \citep[e.g.][]{2014PASP..126..948V,2015ApJ...806...30L,2016MNRAS.459.2408A,2016AJ....152..100L}. For this work we have used two of those pipelines. 

The first one is the EVEREST\footnote{\url{https://stdatu.stsci.edu/prepds/everest/}} (Ecliptic Plane Input Catalog Variability Extraction and Removal for Exoplanet Science Targets) pipeline \citep{2016AJ....152..100L,2018AJ....156...99L} that has shown to provide calibrated data well suited for asteroseismic analyses for red giants \citep[e.g.][]{2017MNRAS.471.2882W,2020ApJS..251...23Z,2022ApJ...926..191Z}. However, for the short cadence data, only light curves up to C13 are available with this calibration. Hence, we also use an adapted version of the Lightkurve Python package\footnote{\url{http://docs.lightkurve.org/}} \citep{2018ascl.soft12013L} to generate the short-cadence light curves for C17. 

\begin{figure}
    \centering
    \includegraphics[width=0.45\textwidth]{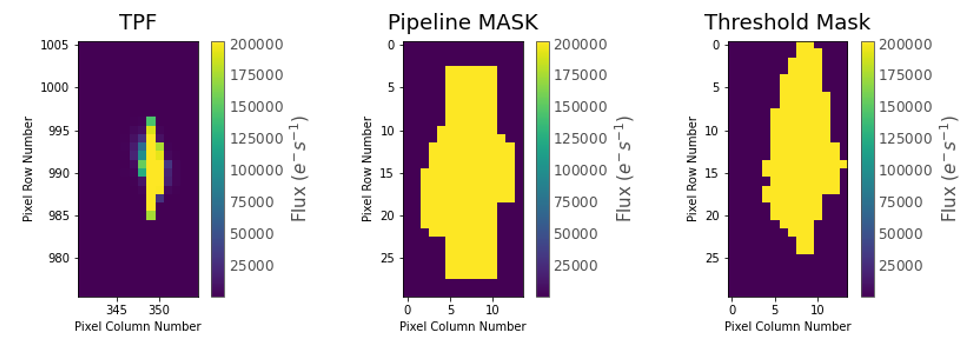}
    \caption{Comparison of the standard K2 aperture (left panel), the standard Lightkurve pipeline mask (middle panel), and our customized mask (right panel) taking pixels with an average flux larger than $1.4$~$10^5$ $e^{-} /s$ for HD~115680 (A). }
    \label{fig:TPF}
\end{figure}

We first extracted the Target Pixel File (TPF) downloaded from the Mikulski Archive for Space Telescopes (MAST). We created a customized aperture for asteroseismic analyses where we took pixels with an average flux larger than 1.4~$10^5$\,$e^{-}/s$.  In Fig.\ref{fig:TPF} we show our customized mask compared to the raw TPF and the pipeline mask for HD~115680 (A).  We then applied the Self Flat Fielding corrector \citep{2018RNAAS...2..182Z}, which takes into account the time scale of the filter as well as the number of chunks in which the 6-hr correction is done. By comparing the Power Spectral Densities (PSD) of the eight stars observed in C6 with both pipelines, we minimized the differences by tweaking the different parameters of the filter. The best results were obtained using 10 segments for the window keyword and a time scale of 1\,day. 


%
\begin{table*}[h!]
\caption{Atmospheric and Global Seismic Parameters for our sample of stars with their corresponding letter references from A to H. } 
\label{table:1} 
\centering 
\resizebox{18.3cm}{!} {
\begin{tabular}{c c c c c c c c c c c c}  
\hline\hline 
Ref & EPIC & HD & Sp. type & $\rm K_{\rm p}$ (mag) & $T_{\rm eff}$ (K) & $L$ (\(\rm L_\odot\)) & [M/H] (dex) & $\nu_{\rm max}$ ($\mu$Hz) & $\Delta \nu$ ($\mu$Hz) & $\log g$ (cgs) & $\epsilon$  \\
\hline 
A & 212478598 & 115680 & K0V & 8.867 & 4925\,$\pm$\,154 & 3.562\,$\pm$\,0.029 & -0.356\,$\pm$\,0.08 & 542\,$\pm$\,23 & 37.60\,$\pm$\,5.33 & 3.65\,$\pm$\,0.06 & 1.43\,$\pm$\,0.03 \\
B & 212485100 & 116832 & F7V & 8.958 & 6081\,$\pm$\,76  & 2.563\,$\pm$\,0.019 & -0.041\,$\pm$\,0.08 & 1874\,$\pm$\,60 & 85.71\,$\pm$\,2.23 & 4.23\,$\pm$\,0.05 & 1.13\,$\pm$\,0.04 \\
C & 212487676 & 114558 & G0V & 9.036 & 6069\,$\pm$\,88  & 2.651\,$\pm$\,0.023 & -0.314\,$\pm$\,0.08 & 1505\,$\pm$\,44 & 77.54\,$\pm$\,2.44 & 4.14\,$\pm$\,0.03 & 1.28\,$\pm$\,0.02 \\
D & 212509747 & 115427 & F5V & 8.310 & 6320\,$\pm$\,128 & 4.487\,$\pm$\,0.037 & -- & 1321\,$\pm$\,59 & 67.61\,$\pm$\,1.88 & 4.09\,$\pm$\,0.05 & 1.05\,$\pm$\,0.04 \\
E & 212516207 & 120746 & F7V & 9.062 & 6024\,$\pm$\,89  & 4.031\,$\pm$\,0.042 & +0.130\,$\pm$\,0.08 & 1248\,$\pm$\,32 & 69.02\,$\pm$\,1.97 & 4.05\,$\pm$\,0.03 & 1.18\,$\pm$\,0.09 \\
F & 212617037 & 117779 & F3IV & 8.720 & 6310\,$\pm$\,259 & 8.150\,$\pm$\,0.159 & -- & 975\,$\pm$\,7 & 49.71\,$\pm$\,2.04 & 3.96\,$\pm$\,0.03 & 1.17\,$\pm$\,0.08 \\
G & 212683142 & 119026 & G1V & 8.914 & 5859\,$\pm$\,61  & 5.670\,$\pm$\,0.052 & -0.052\,$\pm$\,0.08 & 761\,$\pm$\,12 & 45.11\,$\pm$\,2.82 & 3.83\,$\pm$\,0.02 & 1.17\,$\pm$\,0.04 \\
H & 212772187 & 119038 & F5V & 8.948 & 6441\,$\pm$\,139 & 3.103\,$\pm$\,0.024 & -- & 1841\,$\pm$\,87 & 88.72\,$\pm$\,2.43 & 4.24\,$\pm$\,0.06 & 1.01\,$\pm$\,0.01 \\
\hline 
\end{tabular}
}
\flushleft{ Notes. {\it Kepler} magnitude ($\rm K_{\rm p}$) is from the K2 Ecliptic Plane Input Catalog \citep[EPIC, see][]{2016ApJS..224....2H}. $T_{\rm eff}$ and $L$ are taken from {\it Gaia} DR2, [M/H] is from \citep{2021ApJ...922...18O} when available, while $\nu_{\rm max}$, $\Delta \nu$, $\epsilon$ and $\log g$ are derived from our seismic analysis as explained in Section~\ref{sec:4}.}
\end{table*}

Using these two calibration methods, we denote the outputs as: $\rm C6_{\rm EV}$ from EVEREST and $\rm C6_{\rm LK}$ and $\rm C17_{\rm LK}$ from Lightkurve. 
A comparison of the PSDs of HD~115427 (D) is given in Appendix~\ref{appendix:A}. In general, EVEREST light curves have lower noise than those from Lightkurve. 
To increase the frequency resolution of the spectrum, compared to the study of a single, isolated campaign, and improve the overall signal-to-noise ratio of the PSD while having the longer time series to study the oscillation modes and the surface rotation and magnetism, we stitch the data of both campaigns together removing the gap between them. To do so, the first point of C17 is placed just one cadence after the last point of C6. Because these are solar-like pulsating stars with non-coherent modes, the effect of removing the $\sim$2.5-year gap only slightly modifies the widths of the modes that have lifetimes longer than the gap \citep[for more details see][]{2004A&A...423.1051B,2004ESASP.538..265B}. The longest lived p modes in the stars analyzed in this work have lifetimes of around 3.7 days. Therefore, during the 2.5-year gap all the modes are re-excited hundreds of times and thus, there is no effect in the characterization of the central frequencies of the limit spectrum. A detailed Monte-Carlo simulation of the long lived modes in EPIC 212485100 is described in Appendix~\ref{appendix:gap} justifying this choice.


Since $\rm C17_{\rm LK}$ is noisier than $\rm C6_{\rm EV}$, we apply a correction factor to scale the two campaigns on a similar level. To do so, we multiply the $\rm C17_{\rm LK}$ flux by the ratio of the noise in the PSDs above 6\,mHz between C6 and C17 as follows:
\begin{equation}
    \mathrm{flux}'_{\rm C17_{\rm LK}}=\mathrm{flux}_{\rm C17_{\rm LK}}  \frac{ \sqrt{<\mathrm{PSD}_{C6\, >6{\rm mHz}}>}}{\sqrt{<\mathrm{PSD}_{C17\, >{ 6\mathrm{mHz}}}>}} \; ,
 \end{equation}
 where the symbol $<>$ indicates the mean of the PSD.


Light curves from both EVEREST and Lightkurve are then processed with the {\it Kepler} Asteroseismic Data Analysis Correction Software pipeline (KADACS, \citet{2011MNRAS.414L...6G}). Gaps shorter than 5\,days are filled with inpainting techniques using a multi-scale discrete cosine transform \citep{2014A&A...568A..10G,2015A&A...574A..18P} following what has been applied to {\it Kepler} data. The light curves produced in this work are available at gitlab\footnote{\url{https://gitlab.com/rgarcibus/k2_multicampaignk2_gonzalezcuesta2023}}.


\section{Atmospheric parameters}\label{sec:stellparam}


We consolidated the atmospheric parameters (effective temperature and surface gravity) as well as luminosity from the literature in order to use them as inputs in the stellar modeling. 
The most comprehensive catalog for the K2 targets is the Ecliptic Plane Input Catalog \citep[EPIC,][]{2016ApJS..224....2H}, which classified around 138,000 K2 targets providing in particular effective temperature ($T_{\rm eff}$), surface gravity ($\log g$), and metallicity ([Fe/H]) for the K2 stars. Those parameters were inferred from colors, parallaxes, and spectroscopic information when available and the stellar population synthesis model {\it Galaxia} \citep{2011ApJ...730....3S}. However, since the delivery of the EPIC, observations from the {\it Gaia} mission \citep{2005ASPC..338....3P} provide improved and more precise parallaxes with the Data Release 2 \citep{2018A&A...616A...1G}. Thus, we took the DR2 effective temperature and luminosity derived by the {\it Gaia} team. We note that as this work was nearing completion, {\it Gaia} DR3 became available. Some of our targets have spectroscopic parameters in the \texttt{GSP-Spec} module \citep{2022arXiv220605541R}, but a comparison with the DR2 parameters showed that the $T_{\rm eff}$ and $\log g$ values agree within 1 and 2\,$\sigma$. For metallicity, while \texttt{GSP-Spec} values are available for our targets, we did not use them because of their very small error bars, as well as potential systematics in the metallicity scale when compared to other surveys \citep{2022arXiv220605870G}. For five of our targets, \citet{2021ApJ...922...18O} obtained high-resolution spectra providing $T_{\rm eff}$ and [Fe/H]. Their effective temperatures agree with the {\it Gaia} DR2 values within 2\,$\sigma$ so we kept the {\it Gaia} constraints, while we used the metallicity from \citet{2021ApJ...922...18O}. For the remaining three stars, we decided to carry out our analysis using the parameters from {\it Gaia} DR2 that do not include metallicity values (but see Section~\ref{sec:4} for a comparison of the results had we used {\it Gaia} DR3 instead).


\noindent The consolidated atmospheric parameters and {\it Gaia} luminosity of our eight targets are given in Table~\ref{table:1}.



\section{Seismic analysis}\label{sec:4}

Our targets were observed for $\sim$\,160 days when combining the two campaigns. We first estimated the global seismic parameters and then characterized the individual modes.

\subsection{Global Seismic Parameters}

Asteroseismology aims at studying the internal structure and dynamics of the stars by means of their resonant oscillations \citep[e.g.][]{STCDap1993, JCD2002}. These vibrations manifest themselves in small motions of the visible surface of the star and in the associated small variations of stellar luminosity.

In this work, we study the acoustic (p) modes, that are produced in the interior of solar-like stars from the turbulence in their outer layers. P modes of the same degree $\ell$ are asymptotically equidistant in frequency \citep{1980ApJS...43..469T}, which allows us to define some global parameters of the p-mode pattern that we define below.

The first one is the frequency of maximum mode power, $\nu_{\rm max}$. This is the frequency of the maximum of the power envelope of the oscillations (usually fitted by a Gaussian function), where the observed modes present their strongest amplitudes. This parameter can be related to the cut-off frequency in the stellar atmosphere and hence to the stellar surface gravity and the effective temperature of the star \citep{1991ApJ...368..599B,2011A&A...530A.142B}.

The second global seismic parameter is the large frequency spacing, $\Delta \nu$. It is the average spacing in frequency between consecutive radial order modes of the same angular degree. This parameter is directly proportional to the mean density of the star \citep{kjeldsen95}. We determined these two fundamental quantities with two different methods.

\subsubsection{A2Z pipeline}

We performed the first analysis of global seismic parameters with the A2Z pipeline \citep{2010A&A...511A..46M}, where the mean large frequency spacing is computed by considering the power spectrum of the PSD itself. The frequency of maximum power is obtained from fitting a Gaussian on the p-mode bump after having subtracted the background fit on the PSD. The background model consists of 3 components: two Harvey functions \citep{1985ESASP.235..199H} to model different scales of convection and a constant representing the photon noise. Each Harvey function is as follows:

\begin{equation}
    \label{eqn:Harvey_law}
    H(\nu) = \frac{A}{1+\left(\frac{\nu}{\nu_{c}}\right)^\gamma}
\end{equation}

\noindent with A the reduced amplitude component, $\nu_{c}$ the characteristic frequency and $\gamma$ the exponent of the Harvey model that we fixed to 4 \citep[e.g.][]{2014A&A...570A..41K}. Note that for the background fit, we only took into account the PSD above 200\,$\mu$Hz for all the stars in our sample due to the instrumental trends at lower frequencies.

\subsubsection{\texttt{apollinaire} pipeline}

The second method that we applied is the Bayesian Python module \texttt{apollinaire}\footnote{The module documentation is available at \url{https://apollinaire.readthedocs.io/en/latest/} and the source code at \url{https://gitlab.com/sybreton/apollinaire}} from \citet{2022A&A...663A.118B}. Through the implementation of the \texttt{emcee} Ensemble sampler \citep{2013PASP..125..306F}, the module implements Markov Chains Monte Carlo (MCMC) samplings of the parameter posterior probability distribution of background and p-mode models. Following an approach similar to the ones presented in e.g. \citet{2011A&A...525L...9M,2017ApJ...835..172L,2020A&A...640A.130C}, and \citet{2021AJ....161...62N}, the p-mode profile can be fitted with a global parametrization or by considering individual mode parameters as explained in Section~\ref{sec:PB}.


The \texttt{apollinaire} pipeline starts fitting the background based on a model similar to the one of the A2Z pipeline. This model is based on two Harvey functions \citep{1985ESASP.235..199H} with an exponent also fixed to 4, a Gaussian function for the bump of the p modes, and the photon noise. The fit was done with 1000 steps (including 500 steps for the initial burning phase) and 500 walkers, starting at a frequency of 50 $\mu$Hz.

We then fit the asymptotic relation of the p modes based on the \citet{1980ApJS...43..469T} development where the frequency of a mode is described as:
\begin{equation}
   \label{eqn:Tassoul_short}
   \nu_{n,l} \approx  \left(n + \frac{\ell}{2} + \epsilon \right)  \Delta \nu \; ,
\end{equation}
where the order $n$ is much larger than the degree $\ell$ in this approximation and $\epsilon$ is a phase offset. The relation used in \texttt{apollinaire} follows the suggestion of \citet{2017ApJ...835..172L} and extends this relation to the second order
\begin{equation}
    \nu_{n,\ell} \approx \left(n + \frac{\ell}{2}  + \epsilon \right) \Delta \nu - \delta \nu_{0\ell} - \beta_{0\ell} (n - n_\mathrm{max}) + \frac{\alpha}{2} (n - n_\mathrm{max})^2 \; ,
    \label{eq:tassoul_2nd}
\end{equation}
where $\delta \nu_{0 \ell}$ corresponds to the small separations between the modes of degree 0 and the mode of degree $\ell$ \citep[see][for more details]{2022A&A...663A.118B}. $\alpha$ and $\beta_{0\ell}$ are the curvature terms on $\Delta\nu$ and $\delta \nu_{0 \ell}$, respectively. Finally, $n_\mathrm{max}$ is given by
\begin{equation}
    n_\mathrm{max} = \frac{\nu_\mathrm{max}}  {\Delta\nu} - \epsilon \; .
\end{equation}

The description given above is best for main-sequence stars. However for more evolved stars such as subgiants, the presence of mixed modes makes the pattern less regular and a different approach has to be followed. Mixed modes propagate as pressure waves in the convective envelope and as gravity waves in the radiative interior. They typically reach the stellar surface with amplitudes larger than g modes \citep[e.g.][]{2011Sci...332..205B,2011Natur.471..608B,2011A&A...532A..86M}. Therefore, as long as they have enough amplitude at the surface to be detectable, they can be used to probe the inner radiative core with high precision. Unfortunately, the central frequencies of these modes are not simple to be derived with an asymptotic formulation \citep[e.g.][]{2020A&A...642A.226A}. To avoid any perturbation of these mixed modes in \texttt{apollinaire}'s universal pattern fit, the regions of the PSD around these modes are removed and noise is added to have a continuous level between the remaining pairs of modes $\ell =0,2$. A detailed description of the procedure to follow can be found in appendix B of \citet{2022A&A...663A.118B}.


An important parameter of the mode-pattern fit is the phase offset $\epsilon$. Indeed, it allows us to determine whether the identification of the mode degree is correct. This is particularly useful for stars with low signal-to-noise ratio (SNR). To confirm the correct identification we used the $\epsilon$-$T_{\rm eff}$ diagram as shown by \citet{2012ApJ...751L..36W}. $\epsilon$ can be calculated from the fitted frequencies of the $\ell=$0 and the value of $T_{\rm eff}$ by equation \ref{eqn:Tassoul_short}. If the extracted value does not fit the general trend of the $\epsilon-T_{\rm eff}$ relation, it means that the identification is wrong. In that case we re-run the universal pattern fit changing the guess in the $\epsilon$ value given as input.









\subsubsection{Global Seismic Parameters with A2Z and \texttt{apollinaire}}

We ran both codes on the EVEREST calibrated light curves (C6), the Lightkurve data (C6 and C17), and on the stitched C6 and C17 (C6+C17) time series.

We compared the results obtained from the two pipelines for the different observing campaigns ($\rm C6_{\rm EV}$, $\rm C6_{\rm LK}$, $\rm C17_{\rm LK}$) as well as for the combined campaigns (C6+C17). We found that for all the analyzed light curves, the values of $\Delta \nu$ and $\nu_{\rm max}$ obtained with A2Z and \texttt{apollinaire} agree within 3\,$\sigma$, and within 1\,$\sigma$ between $\rm C6_{\rm EV}$ and C6+C17. 

In addition we compared the absolute uncertainties provided by each method. The error bars on $\nu_{\rm max}$ are generally smaller for A2Z by up to a factor of 2, while for $\Delta \nu$, the uncertainties are significantly smaller for \texttt{apollinaire} by a factor of 5 to 6.

Given the results of the comparison between the two pipelines, and to keep the compatibility with previous published values of A2Z for other stars, we decide to use the A2Z global seismic parameters obtained with the combined C6+C17 calibrated light curves for the stellar modeling. We also remind that $\Delta \nu$ values obtained by the A2Z pipeline are ``global''  compared to \texttt{apollinaire}, which performs the universal pattern fit using only 5 orders around $\nu_{\rm max}$, approach usually called ``local'' \citep{2014MNRAS.445.3685C}, which can explain these small differences.

The results obtained for $\nu_{\rm max}$ are combined with $T_{\rm eff}$ from {\it Gaia} DR2 to estimate the seismic $\log g$ using the following equation:

\begin{equation}
    \label{eqn:logg}
    g \simeq g_{\odot} \left( \frac{\nu_{\rm max}}{ \nu_{\rm max,\odot}} \right) \left( \frac{T_{\rm eff}}{ T_{\rm eff,\odot}}\right)^{1/2} \, ,
\end{equation}
where $\nu_{\rm max,\odot}$=3090\,$\pm$\,30\,$\mu$Hz, 
$T_{\rm eff,\odot}= 5777\,$K\footnote{ We note that the IAU has recently adopted a new solar value for the effective temperature where $T_{\rm eff,\odot}$=$5772$\,K \citep[see][]{2016AJ....152...41P}. 
However, since A2Z was calibrated with the previous value of 5777\,K, we need to keep using this temperature.
}, and $g_{\odot}$=27\,402\, $\rm cm \, \rm s^{-2}$ \citep{1991ApJ...368..599B, kjeldsen95}.

Global seismic parameters and surface gravities are given for all the stars in Table~\ref{table:1}.


\subsection{Peak-bagging}\label{sec:PB}

Individual p and mixed modes of all the stars were obtained with the fitting module of \texttt{apollinaire} using 500 walkers and 5500 steps (including a burning phase of 500). From the universal pattern fit described previously, initial guesses of p modes in a range of 7 orders around $\nu_{\rm{max}}$ were automatically created by the code. By visual inspecting these guesses we modify those parameters that seem to be incorrect and perform a first fitting. Then, we divide the PSD by the initial fitted model to obtaining a residual PSD in SNR and all peaks lying where a mode is expected and with a SNR above 8 are added to the guesses and the fit is repeated. Finally, for the two subgiant stars, we do a final fit including the mixed-mode candidates. To do so, we select all peaks with a high SNR that were not identified as $\ell=0$, 2, or 3 modes and we fit them assuming that they could be potential mixed modes. The tables with the resulting frequencies for all the stars are given in Appendix~\ref{appendix:freq} and the \'echelle diagrams are given in Appendix~\ref{appendix:ED}. 

For the five stars in common analyzed by \citet{2021ApJ...922...18O}, the additional modes that we fitted in our work are flagged in the tables of mode frequencies in Appendix~\ref{appendix:freq}. As expected with the longer time series used in this work, we find that in average with our analysis we fit a couple of additional orders at high frequency and one additional order at low frequency as well as higher degree modes ($\ell$\,=2 or 3). We also provide a comparison of the fitted frequencies of the five stars in Appendix~\ref{appendix:comp_Ong}.  We find that the frequencies of the common modes between the two analyses agree within 4\,$\sigma$.

\subsection{Stellar Modelling}\label{sec:model}

Model fitting is based on a set of grids of stellar models. The main one consists of models evolved from the pre-main sequence to the Red Giant Branch (RGB) using the MESA code \citep[Modules for Experiments in Stellar Astrophysics,][]{2011ApJS..192....3P,2013ApJS..208....4P,2015ApJS..220...15P}, version 15$\,$140.
The OPAL opacities \citep{1996ApJ...464..943I} and the GS98 metallicity mixture \citep{1998SSRv...85..161G} were used,
otherwise the standard input physics from MESA was applied. 
The grid is composed of evolutionary sequences with masses $M$ from $0.8\rm M_{\odot}$ to $1.5\rm M_{\odot}$ with a step of $\Delta M = 0.01\rm M_{\odot}$, initial abundances [M/H] from  $-0.3$ to  $0.4$ with a step of $0.05$, and mixing length parameters $\alpha$ from $1.5$ to $2.2$ with a step of $\Delta \alpha=0.05$.  The mixing length theory is modeled according to \citet{1968pss..book.....C}. Eigenfrequencies were computed in the adiabatic approximation using the ADIPLS code \cite{2008Ap&SS.316..113C}.

A second grid of models was built with the same range of parameters but including microscopic diffusion. A third grid includes diffusion and overshooting, implemented with the exponential prescription given by \cite{2000A&A...360..952H}. This grid was limited to masses between $1.3\rm M_{\odot}$ and $1.39\rm M_{\odot}$ and the overshooting parameter was fixed to $f=0.02$, which is a standard value in this description  \citep[see][for a quick reference to the equation]{2019FrASS...6...41P}.

The initial metallicity $Z$ and helium abundance $Y$ were derived from [M/H], constrained by taking a Galactic chemical evolution model with $\Delta Y/\Delta Z= (\rm Y_{\odot} - Y_0) / \rm Z_{\odot}$ fixed. Assuming a primordial helium 
abundance of $Y_0=0.249$ and initial solar values of $Y_{\odot}=0.2744$ and $\rm Z_{\odot}=0.0191$ (consistent with the opacities and GS98 abundances  considered above) a value of $\Delta Y/\Delta Z= 1.33$ is obtained. A surface solar metallicity of $\rm (Z/X)_{\odot}=0.0229$ was used to derive values of $Z$ and $Y$ from the [M/H] interval. 
Hereafter we refer to this set of model grids as IACgrid.

For a typical evolutionary sequence in the initial grid, 
we save about 100 models from the zero age main sequence (ZAMS) to the terminal age main sequence (TAMS). Owing to the very rapid change in the dynamical time scale of the models, $t_{\text{dyn}}=(R^3/GM)^{1/2}$, such grid is too coarse in the time steps. Nevertheless, the dimensionless frequencies of p modes change so slowly that interpolations between models introduce errors much lower than the observational ones. This procedure was discussed in more detail in \cite{2016A&A...591A..99P} and was found safe and consumes relatively less time.

The stellar parameters are found through a $\chi^2$ minimization that compares observed values to the grid of models discussed above. The general procedure is similar to that described in \citet{2019FrASS...6...41P}. Specifically we minimize the function
\begin{equation}
\chi^2=\frac{1}{3} \left( \chi^2_{\mathrm{freq}} +\chi^2_{\mathrm{spec}} +\chi^2_{\mathrm{dyn}} \right)
\; .
\label{eq_chi2}
\end{equation}

Here, we define:
\begin{equation}
\chi^2_{\mathrm{spec}}= \frac{1}{3} \left[ \left( \frac{\delta T_{\text{eff}}}{\sigma_{T_{\text{eff}}}} \right)^2+ 
\left(\frac{\delta g}{\sigma_{g}}\right)^2 + \left(\frac{\delta (Z/X)}{\sigma_{ZX}}\right)^2 + 
\left(\frac{\delta L}{\sigma_{L}}\right)^2
\right]
\; ,
\end{equation}
where $\delta T_{\text{teff}}$, $\delta g$, $\delta (Z/X)$ and $\delta L$ correspond to differences between the observations and the 
models whereas $\sigma_{T_{\text{eff}}}$, $\sigma_{g}$, $\sigma_{Z/X}$ and $\sigma_{L}$ are their respective observational errors.
Values for $L$, $T_{\text{eff}}$ and their errors are given in Table~\ref{table:1} but $\log g$ was derived from $\nu_{\text{max}}$ and $T_{\text{eff}}$ using the seismic scaling relation (see Eq.~\ref{eqn:logg}). Here a systematic error of $0.1$\,dex for the main sequence stars and $0.2$\,dex for the red giant was assumed. Finally, as mentioned before, values of $Z/X$ where taken from 
\citet{2021ApJ...922...18O} when available.


The term $\chi^2_{\text{freq}}$ in Eq.~\eqref{eq_chi2} corresponds to the frequency differences between the models and the observations after removing a smooth function of frequency in order to filter out surface effects not considered in the modelling. This surface term is computed only using radial oscillations as described in \citet{2019FrASS...6...41P}.
When the surface term is determined, we consider radial as well as non-radial modes for computing the corresponding minimization function $\chi^2_{\rm freq}$.

Finally, $\chi^2_{\rm dyn}$ takes into account the dynamical time. For more information on the $\chi^2$, we refer to \cite{2016A&A...591A..99P} and \cite{2019FrASS...6...41P}.

To estimate the uncertainty in the output parameters we assumed normally distributed uncertainties for the observed frequencies, mean density, and spectroscopic parameters. We then search for the model with the minimum $\chi^2$ in every realization and compute mean values and standard deviations.




\section{Discussion}
\label{sec:5}

\subsection{Comparison between the observed and predicted frequency of maximum power}

The seismic scaling relation given in Eq.~\ref{eqn:logg} can be inverted to calculate an estimation of the frequency of maximum power if we have the effective temperature and the surface gravity of a star. Therefore, thanks to the EPIC, we obtain a first estimation of the global seismic parameters.

A second estimation of $\nu_{\rm max}$ can be calculated using the effective temperatures and radii from the {\it Gaia} DR2 stellar parameters catalog \citep{2018A&A...616A...8A} using Eq. ~\ref{eqn:NuMax_Gaia}:
\begin{equation}
    \label{eqn:NuMax_Gaia}
    \nu_{\rm max} = \nu_{\rm max,\odot} \left(\frac{R}{R_{\odot}}\right)^{-1.85} 
    \left(\frac{T_{\rm eff}}{T_{\rm eff,\odot}}\right)^{0.92} ,
\end{equation}

\noindent where $\nu_{\rm max,\odot}$=3090\,$\pm$\,30\,$\mu$Hz, $T_{\rm eff,\odot}$=5777\,K and $R_{\odot}$=6.957e+10\,cm (\citet{2013ARA&A..51..353C}; \citet{2013ASPC..479...61B}).




\begin{figure}[h!]
    \centering
    \includegraphics[width=0.5\textwidth]{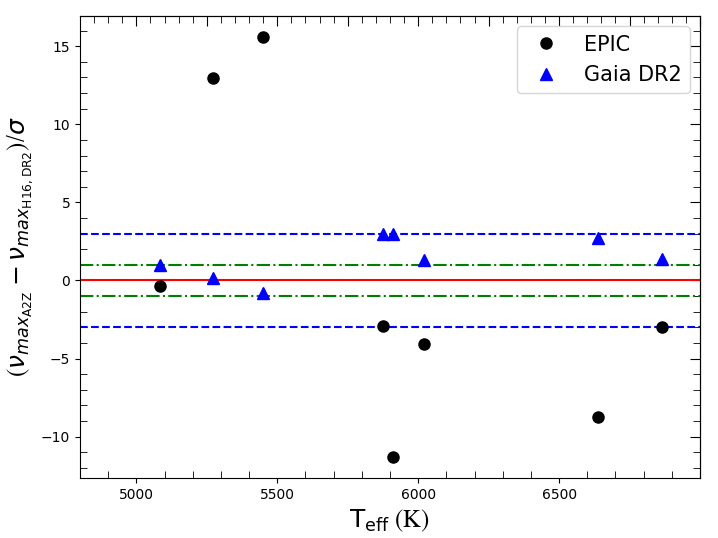}
    \caption{Ratio of the difference between the observed $\nu_{\rm max}$ and the estimation from EPIC data (black circles) and Gaia DR2 data (blue triangles), and the combined uncertainties, $\sigma$. The green dot-dash lines correspond to $\pm$\,1\,$\sigma$ and the blue dash lines represent the 3\,$\sigma$ limits, where sigma is the square root of the sum of the quadratic errors. The red continuous line depicts the null difference.}
    \label{NuMax_C6_17_A2Z_H16_DR2_Teff}
\end{figure}

Figure \ref{NuMax_C6_17_A2Z_H16_DR2_Teff} shows the comparison between the observed $\nu_{\rm max}$ and the expected values from EPIC and {\it Gaia} stellar parameters. We can see some difference between the two catalogs. Indeed, the effective temperatures are in many cases very different between the EPIC and {\it Gaia}. As expected, estimations of $\nu_{\rm max}$ using data from {\it Gaia} DR2 are more consistent with the seismic observations than those obtained from the EPIC data as for the majority of the stars they agree within 1\,$\sigma$.
This suggests that the precision of the {\it Gaia} DR2 parameters is enough to predict the frequency region of the p modes for our sample of solar-like stars.





\subsection{Stellar Parameters from asteroseismology}

\label{sec:res_stellparam}

Using $T_{\rm eff}$, $L$, and  $\nu_{\rm max}$ combined with the individual frequencies of the p modes, we looked for the best-fit models of our eight targets with the IACgrid described in Section~\ref{sec:model}. The main stellar parameters from the best-fit model are obtained with the grid without treatment of diffusion. They are listed in Table~\ref{table:2}. We also list the results coming from the grid that includes diffusion, which allows us to have a sense of the effect of different physics in the models. We will come back to this point later.

\begin{figure}[h]
    \centering
    \includegraphics[width=10cm]{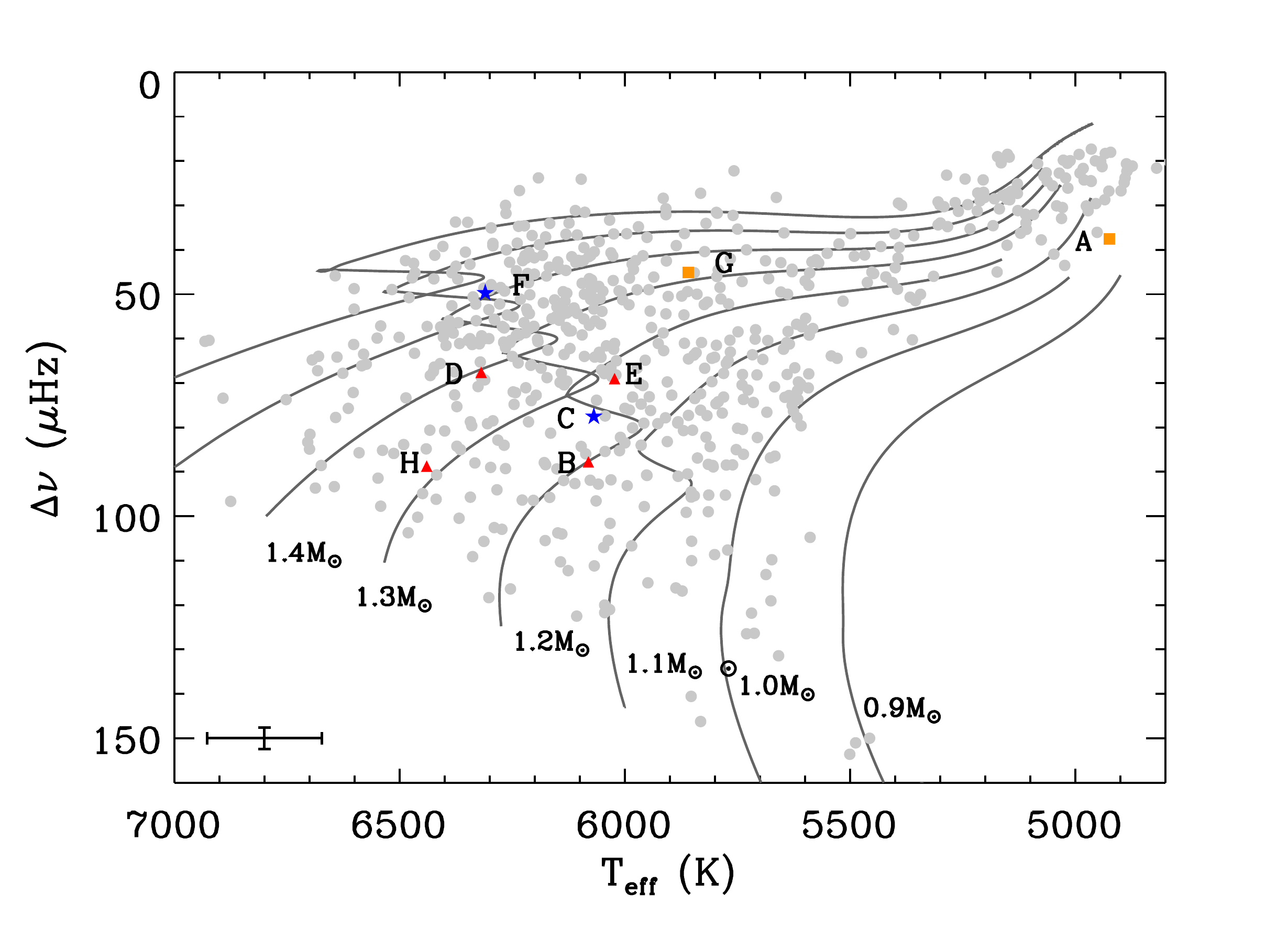}
    \caption{Seismic Hertzsprung-Russell Diagram where the luminosity is modified by the mean large frequency spacing, $\Delta \nu$. The colored symbols represent the stars studied here: red triangles are the main-sequence stars; blue stars are stars near the TAMS; orange squares are subgiants. The {\it Kepler} sample with detected solar-like oscillations \citep{2022A&A...657A..31M} are also represented with grey circles. The evolutionary tracks with solar metallicity from the Aarhus STellar Evolution Code \citep{2008Ap&SS.316...13C} are represented with the black lines. The average uncertainties on $T_{\rm eff}$ and $\Delta \nu$ are shown in the lower left corner.}
    \label{HRD}
\end{figure}

In Figure~\ref{HRD}, we show the location of our targets in a modified seismic Hertzsprung-Russell diagram (HRD) along with the 624 {\it Kepler} solar-like stars with a detection of p modes \citep{2022A&A...657A..31M}. Our sample of stars is in general more massive than the Sun (above 1.2\,$\rm M_\odot$). We can also see the two evolved stars: one subgiant (HD~119026 (G)) and one early red giant (HD~115680 (A)). In addition, we note that from the best-fit models, two stars HD~114558 (C) and HD~117779 (F) are close to the Terminal Age Main Sequence (TAMS). 



In the last column of Table~\ref{table:2}, we list the $\chi^2$ values for the best-fit models. In general, all well determined modes were used in the fit but if this implied values $\chi^2>3$ then mixed $\ell=1$ modes were not considered. In general, this happens for the subgiants. Our IACgrid has only four free parameters: mass, metallicity, mixing length parameter, and age while we are fitting four observed parameters and the individual mode frequencies. Furthermore, the frequency corrections considered to take into account surface uncertainties include mode inertia but not effects of the coupling between $\ell=1$ mixed modes. Although mixed modes can be very useful to probe the physics of the stellar interior, providing stellar parameters (such as age) with smaller uncertainties, the inclusion of these modes will yield results that are more model dependent than the ones reported here. 

To better illustrate the fit of the observed frequencies, in Appendix~\ref{appendix:ED} we show  \'echelle diagrams of the eight stars with the comparison between the model frequencies with diffusion and the observed ones. 



Looking at the \'echelle diagrams in Appendix~\ref{appendix:ED}, we also note that HD~119038 (H) and HD~115427 (D) have lower SNR compared to the other stars. 

For HD~117779 (F), the {\it Gaia} data provide large error bars on both the luminosity (5 times larger than for the other stars) and the effective temperature (250\,K). The luminosity value also seems to be quite high given the seismic parameters, which makes it more dubious. The fit without the luminosity constraint leads to a lower $\chi^2$ value (going from 2.1 to 1.1).  We can also see in the \'echelle diagram that the ridges are not so clear. Being an evolved F-dwarf, the width of the modes is larger, which could lead to some confusion on the identification of the modes. Comparing the fitted frequencies to the model frequencies, the dipolar modes with $n$\,=\,12 and 13 were not used for the modeling because at this frequency range, the presence of mixed modes in the models bump the p-dominated modes and we are not able to stabilize the minimization when those modes are used.




\begin{figure}[h!]
    \centering
    \includegraphics[width=9cm]{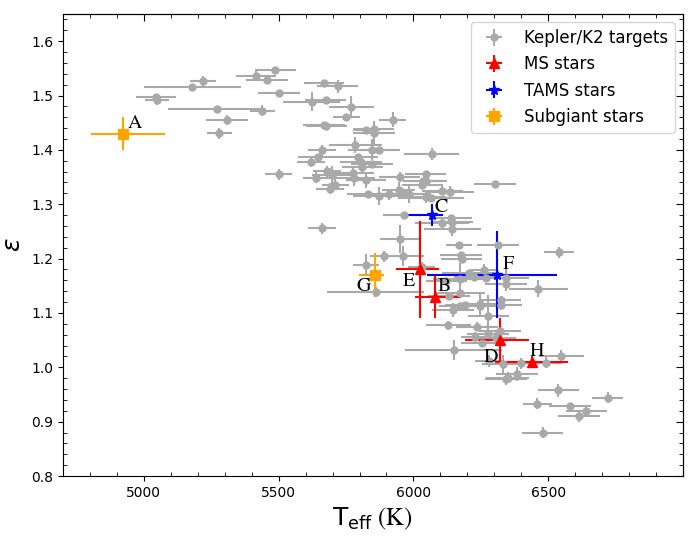}
    \caption{Phase offset from the universal pattern fit of \texttt{apollinaire}, $\epsilon$, as a function of $\rm T_{\rm eff}$ from {\it Gaia} DR2. The {\it Kepler} targets with characterized individual modes are represented with grey circles. MS stars in our sample, the two TAMS stars and the two subgiants are represented with red triangles, blue stars and orange squares, respectively.}
    \label{epsilon_plot}
\end{figure}

As mentioned above, the identification of the modes can be done with the phase offset $\epsilon$ in Eq.~\ref{eqn:Tassoul_short}. From the universal pattern of \texttt{apollinaire}, we thus obtained an estimation for that parameter and represent it in the $\epsilon-T_{\rm eff}$ diagram (see Figure~\ref{epsilon_plot}). For comparison we also added a sample of 119 \emph{Kepler} solar-like stars where the frequencies of the individual modes were obtained by \citet{2012A&A...543A..54A}, \citet{2016MNRAS.456.2183D}, and \citet{2017ApJ...835..172L} (grey circles). We can see the correlation between the two parameters: as the effective temperature increases, the $\epsilon$ value decreases. This emphasizes that the identification of the individual frequencies of our eight targets studied is the correct one. The early red giant, HD~115680 (A), is the coolest star in this diagram seating completely in the left hand side.

\begin{figure*}[h!]
    \centering
    \includegraphics[width=17cm]{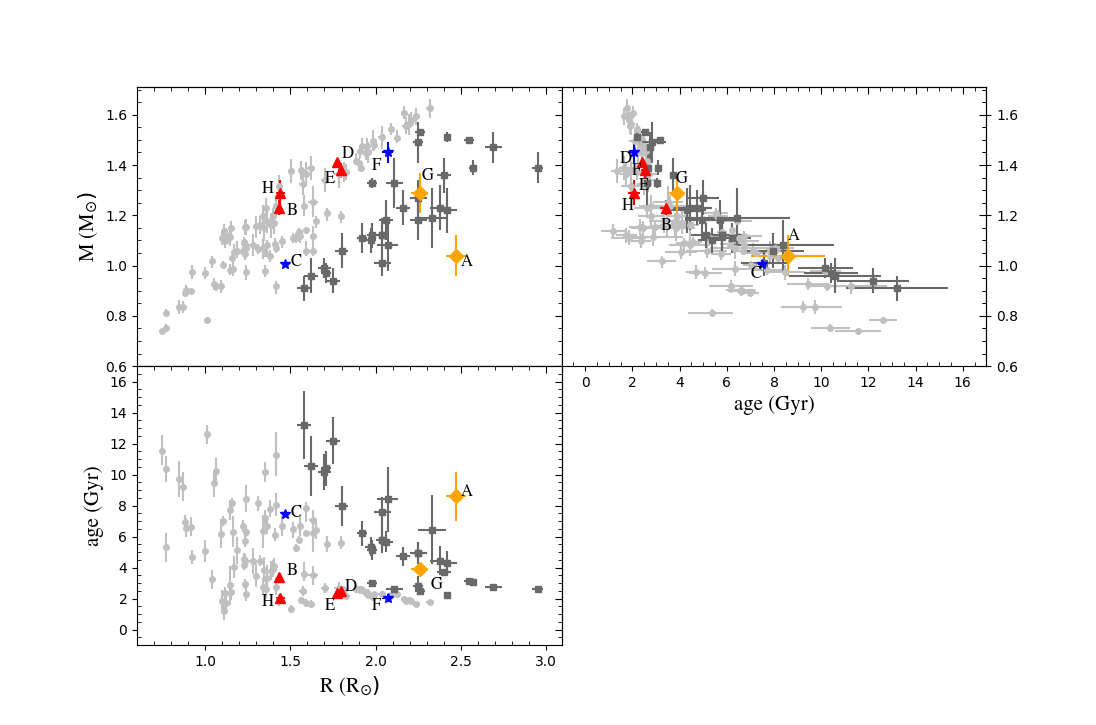}
    \caption{Stellar fundamental parameters obtained with models without diffusion for our K2 targets (referred to through with letters A through H) showing mass vs radius (upper left panel), mass vs age (upper right panel), and age vs radius (lower left panel). In all panels, main-sequence stars are represented by red triangles, stars close to the TAMS with blue stars and subgiant/red giant with orange diamonds. For comparison, {\it Kepler} solar-like stars with detailed seismic modeling are represented with grey circles \citep{2015MNRAS.452.2127S,2017ApJ...835..173S} while subgiants are shown with dark grey squares \citep{2020MNRAS.495.3431L}.}
    \label{M_R_plot_Silva_1517_Li20}
\end{figure*}

In Figure~\ref{M_R_plot_Silva_1517_Li20}, we compare the stellar parameters of our sample with the {\it Kepler} solar-like stars with detailed modeling from \citet{2015MNRAS.452.2127S,2017ApJ...835..173S}, represented with grey dots, and the subgiant sample modeled by \citep{2020MNRAS.495.3431L} with dark grey squares. The mass as a function of the radius is shown in the upper left panel. The top right panel represents the mass as a function age and, finally, in the bottom panel, we show the age as a function of radius. 



While the main-sequence stars populate the same region of the M-R diagram, the more evolved stars move towards larger radius, as expected by stellar evolution. The two stars close to the TAMS (HD~114558 (C) and HD~117779 (F)) fall in the less populated region between the main-sequence and subgiant stars. 


It is also interesting to note that an evolved 1\,$\rm M_{\odot}$ is in our sample, in a region of the diagram not very populated. This is the early red giant in our sample, HD~115680 (A) star, with a radius of 2.47\,$\pm$\,0.06 $\rm R_{\odot}$. 

We can also see in the top right panel of Figure~\ref{M_R_plot_Silva_1517_Li20} that for a given mass, the evolved stars from our sample are among the oldest already known from the {\it Kepler} main-mission sample.



\begin{figure}[h!]
    \centering
    \includegraphics[width=8cm]{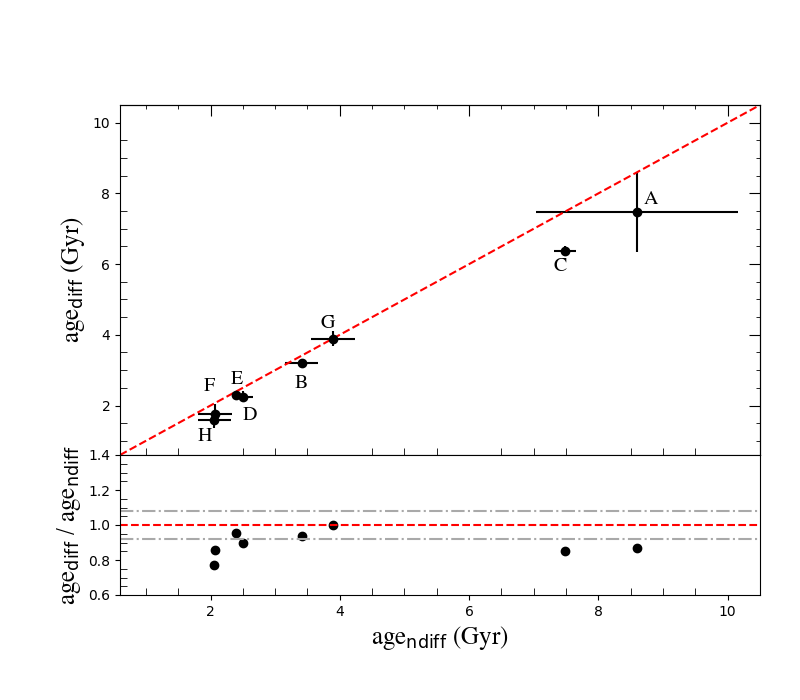}
    \caption{Ages obtained with the IACgrid with diffusion versus ages from the IACgrid without diffusion (top panel). The one to one line is represented with the red dash line. Ratio between the two different age computations as a function of the age from the IACgrid without diffusion (bottom panel). The red dash line is the equality line while the grey dot-dash lines correspond to the mean age uncertainty on stellar ages without diffusion of our sample (of 8\%).}
    \label{age_dif_ndif_plot}
\end{figure}

We note that in general the changes in the stellar parameters when including diffusion are inside the internal uncertainties except for the ages. Indeed, microscopic diffusion impacts the abundance of chemical elements inside the stars as they can migrate from the surface to the interior. As a consequence, it can change the amount of H in the core and modify the stellar ages (Fig.~\ref{age_dif_ndif_plot}). Models without diffusion are older than models with diffusion. In the bottom panel of Figure~\ref{age_dif_ndif_plot}, we can see that the differences between the ages are larger than 10\% in most of the stars. The mean internal error on the ages of 8\% is clearly underestimating the systematic uncertainty in most of the cases.


For one star, HD~119038 (H), with a mass $\rm M\simeq 1.3\,\rm M_\odot$, we consider a fit to a grid of models with overshooting, as indicated in Section~\ref{sec:model}. As can be seen in Table~\ref{table:2}, stellar parameters from the grid with overshooting are consistent with those derived from the grid without including it.

Finally we compute results considering 
$T_\mathrm{eff}$ and $\log g$ from {\it Gaia} DR3 database as mentioned in Section~\ref{sec:stellparam}. The ages, masses, and radii change on average by $13\%$, $6\%$,  and $3\%$ respectively. These figures are slightly higher than the statistical errors given in Table~\ref{table:1}. As mentioned previously, these results should be taken with caution since the spectroscopic values of $\log g$ in DR3 require further analysis.

\begin{table*}[h]
\caption{Stellar parameters from the seismic modeling. The lines with ``no $L$'' means that the Luminosity derived from {\it Gaia} were not used in the fit. The column with $Z/X$ corresponds to surface values.} 
\label{table:2} 
\centering 
\resizebox{18.3cm}{!} {
\begin{tabular}{c c c c c c c c c c c}  
\hline\hline 
EPIC & Ref & Diffusion & $M$ (\(\rm M_\odot\)) & $R$ (\(\rm R_\odot\)) & $\rho$ (cgs) & $\log$ g (dex) & $T_{\rm eff}$ (K) & Age (Gyr) & $Z/X$ & $\chi^2$\\ 
\hline 
212478598 & A & No & 1.04\,$\pm$\,0.08 & 2.47\,$\pm$\,0.06 & 0.0968\,$\pm$\,0.0003 & 3.667\,$\pm$\,0.011 & 5030\,$\pm$\,60 & 8.60\,$\pm$\,1.56 & 0.012\,$\pm$\,0.005 & 2.9 \\
& & Yes & 1.04\,$\pm$\,0.04 & 2.48\,$\pm$\,0.03 & 0.0967\,$\pm$\,0.0003 & 3.666\,$\pm$\,0.004 & 5042\,$\pm$\,42 & 7.47\,$\pm$\,1.13 & 0.019\,$\pm$\,0.003 & 5.4 \\
212485100 & B & No & 1.23\,$\pm$\,0.03 & 1.433\,$\pm$\,0.014 & 0.599\,$\pm$\,0.007 & 4.215\,$\pm$\,0.003 & 6099\,$\pm$\,35 & 3.41\,$\pm$\,0.26 & 0.011\,$\pm$\,0.014 & 1.6 \\
& & Yes & 1.19\,$\pm$\,0.02 & 1.412\,$\pm$\,0.011 & 0.611\,$\pm$\,0.008 & 4.2140\,$\pm$\,0.0013 & 6145\,$\pm$\,21 & 3.19\,$\pm$\,0.12 & 0.023\,$\pm$\,0.003 & 1.2 \\
212487676 & C & No & 1.005\,$\pm$\,0.009 & 1.468\,$\pm$\,0.005 & 0.450\,$\pm$\,0.004 & 4.1060\,$\pm$\,0.0013 & 6075\,$\pm$\,24 & 7.48\,$\pm$\,0.17 & 0.012\,$\pm$\,0.002 & 1.8 \\
& & Yes & 1.025\,$\pm$\,0.008 & 1.470\,$\pm$\,0.004 & 0.455\,$\pm$\,0.005 & 4.1136\,$\pm$\,0.0013 & 6068\,$\pm$\,15 & 6.37\,$\pm$\,0.14 & 0.011\,$\pm$\,0.001 & 1.3 \\
212509747 & D & No & 1.38\,$\pm$\,0.03 & 1.794\,$\pm$\,0.017 & 0.351\,$\pm$\,0.003 & 4.070\,$\pm$\,0.002 & 6272\,$\pm$\,29 & 2.51\,$\pm$\,0.14 & 0.031\,$\pm$\,0.006 & 1.0 \\
& & Yes & 1.39\,$\pm$\,0.05 & 1.787\,$\pm$\,0.025 & 0.357\,$\pm$\,0.006 & 4.076\,$\pm$\,0.003 & 6286\,$\pm$\,42 & 2.25\,$\pm$\,0.15 & 0.0030\,$\pm$\,0.007 & 1.0 \\
212516207 & E & No & 1.414\,$\pm$\,0.010 & 1.775\,$\pm$\,0.006 & 0.367\,$\pm$\,0.003 & 4.0894\,$\pm$\,0.0011 & 6138\,$\pm$\,25 & 2.40\,$\pm$\,0.05 & 0.005\,$\pm$\,0.015 & 1.9 \\
& & Yes & 1.37\,$\pm$\,0.03 & 1.743\,$\pm$\,0.015 & 0.374\,$\pm$\,0.006 & 4.0915\,$\pm$\,0.0020 & 6195\,$\pm$\,33 & 2.29\,$\pm$\,0.01 & 0.032\,$\pm$\,0.007 & 1.6 \\
212617037 & F & No & 1.45\,$\pm$\,0.04 & 2.07\,$\pm$\,0.03 & 0.232\,$\pm$\,0.005 & 3.967\,$\pm$\,0.005 & 6748\,$\pm$\,65 & 2.07\,$\pm$\,0.26 & 0.033\,$\pm$\,0.011 & 1.9 \\
& & Yes & 1.46\,$\pm$\,0.04 & 2.063\,$\pm$\,0.020 & 0.235\,$\pm$\,0.004 & 3.972\,$\pm$\,0.005 & 6746\,$\pm$\,43 & 1.77\,$\pm$\,0.26 & 0.016\,$\pm$\,0.002 & 2.1 \\
& & No, no $L$ & 1.40\,$\pm$ 0.05 & 2.08\,$\pm$ 0.05 &  0.228\,$\pm$ 0.002 &  3.947\,$\pm$ 0.009 & 6410\,$\pm$ 210 &  2.65\,$\pm$ 0.24 & 0.029\, $\pm$  0.028 & 1.1 \\
& & Yes, no $L$ & 1.36\,$\pm$ 0.05 & 2.07\,$\pm$ 0.06 &  0.223\,$\pm$ 0.008 &  3.939\,$\pm$ 0.013 & 6318\,$\pm$ 287 &  2.9\,$\pm$ 0.3 & 0.031\, $\pm$  0.016 & 1.1 \\
212683142 & G & No & 1.29\,$\pm$\,0.08 & 2.26\,$\pm$\,0.05 & 0.159\,$\pm$\,0.002 & 3.840\,$\pm$\,0.007 & 5920\,$\pm$\,64 & 3.90\,$\pm$\,0.34 & 0.012\,$\pm$\,0.011 & 1.0 \\
& & Yes & 1.26\,$\pm$\,0.06 & 2.24\,$\pm$\,0.04 & 0.159\,$\pm$\,0.002 & 3.838\,$\pm$\,0.005 & 5949\,$\pm$\,52 & 3.89\,$\pm$\,0.21 & 0.022\,$\pm$\,0.006 & 1.0 \\
212772187 & H & No & 1.29\,$\pm$\,0.05 & 1.441\,$\pm$\,0.025 & 0.636\,$\pm$\,0.015 & 4.232\,$\pm$\,0.003 & 6384\,$\pm$\,58 & 2.06\,$\pm$\,0.26 & 0.033\,$\pm$\,0.011 & 1.0 \\
& & Yes & 1.30\,$\pm$\,0.04 & 1.435\,$\pm$\,0.020 & 0.651\,$\pm$\,0.016 & 4.238\,$\pm$\,0.002 & 6394\,$\pm$\,45 & 1.59\,$\pm$\,0.23 & 0.030\,$\pm$\,0.010 & 1.3 \\
& & Over & 1.33\,$\pm$\,0.02 & 1.442\,$\pm$\,0.012 & 0.645\,$\pm$\,0.014 & 4.244\,$\pm$\,0.001 & 6373\,$\pm$\,32 & 1.20\,$\pm$\,0.11 & 0.023\,$\pm$\,0.003 & 2.1 \\
\hline 
\end{tabular}
}
\end{table*}

\subsection{Comparison between {\it Gaia} DR2 and seismic radii}

\begin{figure}[h]
    \centering
    \includegraphics[width=8cm]{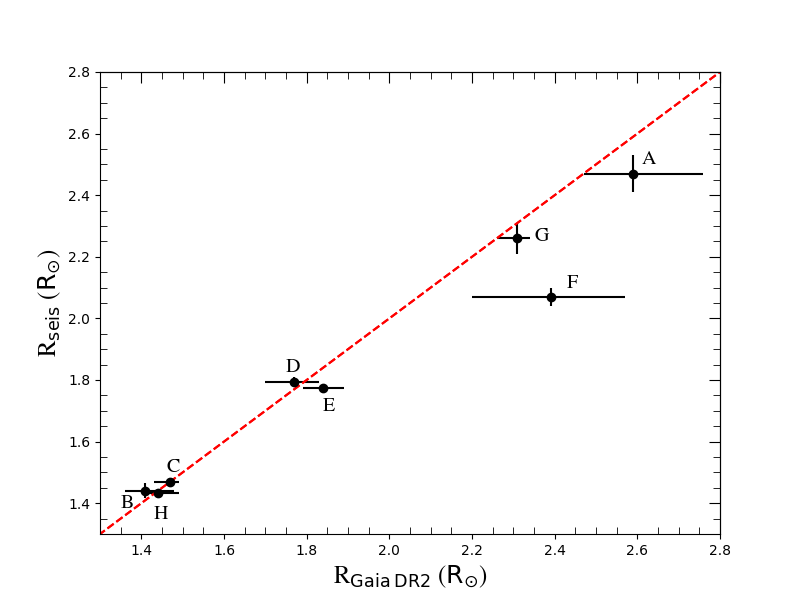}
    \caption{Comparison between {\it Gaia} DR2 radii and radii from seismic modeling without using the luminosity constraint from {\it Gaia} for our eight targets. The red dashed line represents the one to one line.}
    \label{comp_R}
\end{figure}

The {\it Gaia} mission has provided radii for a large number of stars in our galaxy, including our eight targets. We decide here to compare the {\it Gaia} radii with the seismic ones. However, because we used the luminosity as an input in the models, the seismic analysis is clearly biased towards the {\it Gaia} radii. For an independent comparison, we also computed the best-fit models without the luminosity constraint. The result of the comparison is shown in Figure~\ref{comp_R}. The agreement between {\it Gaia} and asteroseismic radii is in general quite good for all the targets except for HD~117779 (F), which is 1.5\,$\sigma$ away. As said above, this star has a large uncertainty on the luminosity and effective temperature as well as a low SNR in the PSD, making the identification of the modes more complicated. As a consequence, the seismic analysis of that star should be taken with caution. 


On average, we find that the difference between the seismic and {\it Gaia} radii agree is of 4\% with a scatter of 5\%. The seismic radii are slightly underestimated. However the disagreement is larger for more evolved stars in the subgiant phase. This agrees with the previous comparisons between {\it Gaia} and asteroseismology done for hundreds of solar-like stars observed by the {\it Kepler} mission \citep[e.g.][]{2017ApJ...844..102H,2019ApJ...885..166Z,2022A&A...657A..31M}.

\subsection{Study of the surface rotation}

Stellar rotation periods are useful not only for the study of angular momentum transport in stars \citep{2020A&A...636A..76S} but also to infer stellar ages via gyrochronology \citep{2007ApJ...669.1167B} given that stars on the main sequence were shown to spin down from spectroscopic observations of young clusters \citep{1972ApJ...171..565S}. Rotation periods, obtained from asteroseismology or modulations in the light curve related to the presence of active regions and/or faculae at the surface of the star, were combined with seismic ages, which showed that some stars rotate faster what was expected from classical gyrochronology relations \citep{2015MNRAS.450.1787A,2016Natur.529..181V,2021NatAs...5..707H}. This suggested that the magnetic braking weakens when stars go beyond a certain point in their evolution. Although most of the stars in our sample may be too evolved or too hot for a Skumanich-type spin-down to take place, it is nonetheless interesting to compare the gyrochronology predictions with analysis from the empirical light curves. 

The estimation of surface rotation has been done in large samples of stars observed by {\it Kepler} and K2 \citep[e.g.][]{2013A&A...557L..10N,2014A&A...572A..34G,2014ApJS..211...24M,2019ApJS..244...21S,2020A&A...635A..43R,2021ApJ...913...70G,2021ApJS..255...17S}. This measurement relies on the passage of active regions and/or faculae on the visible stellar disk with a periodicity related to the rotation of the star.

To study the surface rotation of our targets, we only use the C6 EVEREST light curves as the Lightkurve data (the only ones available for C17) are heavily filtered at low frequency. We searched for the presence of a modulation in the light curves by combining a time-frequency analysis based on wavelets \citep{1998BAMS...79...61T,liu2007,2010A&A...511A..46M} and the auto-correlation function \citep[ACF;][]{2014A&A...572A..34G,2014ApJS..211...24M}. Given the length of the data of 80 days, we should be able to determine reliable periods up to around 20 days.



Only two stars present some clear modulation in the analysis. For HD~120746 (E), which is 40\% more massive than the Sun and with a seismic age of 2.5\,Gyr, a periodicity around 28-31\,days appears in both the wavelet power spectrum (WPS) and the ACF. Unfortunaltely, this value is above the aforementioned limit of 20 days and thus, the retrieved periodicity could be a lower limit of the real rotation period.


The second star, HD~117779 (F), has a mass of 1.4\,$\rm M_\odot$ and  is close to be a subgiant. Our analysis shows different fast possible rotation periods of 1.9\,days, 4.4\,days, and 9.6\,days. For such a massive star, the magnetic braking is weaker than a typical low-mass solar-like star below the Kraft break \citep{1967ApJ...150..551K}, and hence, it can retain a fast rotation even at this late stage of the main sequence. Longer datasets would be necessary to confirm its rotation period.

\subsection{Looking for frequency shifts induced by an underlying magnetic cycle}\label{sec:freqshift}

For the Sun, it has been observed that its surface magnetic activity affects the properties of the acoustic modes: when the magnetic activity increases the frequencies of the modes are shifted towards higher frequencies, the widths are increased, and the amplitude of the modes decrease \citep[e.g.][]{1990Natur.345..322E,1992A&A...255..363A}. This behaviour has also been observed in other solar-like stars with space missions like CoRoT and {\it Kepler} \citep[e.g.][]{2010Sci...329.1032G,2016A&A...589A.118S,2017A&A...598A..77K,2018ApJS..237...17S, 2019FrASS...6...46M}. 

K2 multi-campaign observations of the same solar-like pulsating stars separated by nearly three years open the possibility to look for changes in the p-mode parameters that could be related to changes in magnetic activity that could uncover the presence of magnetic cycles in these stars.

Following \citet{2010Sci...329.1032G}, we analyzed subseries of 26-day long shifted by 13 days, yielding 5 subseries for C6 (calibrated with EVEREST pipeline) and 4 subseries for C17 (using Lightkurve) for the main-sequence solar-like stars. HD~115680 (A) and HD~119026 (G), which are subgiants, were not analyzed as we do not expect them to be very active at this evolutionary stage. Using the same \texttt{apollinaire} setup as the one described for the fitting of the modes in Section~\ref{sec:PB}, we extracted the properties of the modes for each subserie. Frequency shifts were computed using as a reference those obtained in the first subserie and the weighted means and standard deviations of modes $\ell$=0, 1, and 2 were calculated. 

Unfortunately, the frequency shifts obtained are compatible with no variation at one sigma level. The average uncertainties are in the range $\sim$0.2 to $\sim$0.4 $\mu$Hz, which is of the order of the maximum variation of the frequency shifts observed in the Sun and half of the maximum frequency shifts observed in other \emph{Kepler} stars by \citet{2018ApJ...852...46K,2018ApJS..237...17S}. To improve the quality of the fits and reduce the uncertainties, we have also compared directly the frequencies of C17 with the ones for C6 (used as a reference) and, again, no variation is observed within the uncertainties. Thus, we can consider an upper limit of $\sim$0.4 $\mu$Hz for the average frequency shift in these stars. 

Higher SNR data and more modes at high frequency (those more sensitive to the magnetic perturbations) would be required to unambiguously detect magnetic-activity related changes in the properties of p modes in these stars.


\section{Conclusions}

We have presented the asteroseismic analysis of eight solar-like stars using photometric data of the C6 and C17 observation campaigns of the K2 mission. By concatenating the EVEREST C6 light curve and the C17 light curve produced by an adapted version of the Lightkurve package, we obtain light curves of $\sim$160 days observation length.

By analyzing the K2 data with the two seismic pipelines, A2Z and \texttt{apollinaire}, we characterize the solar-like oscillations with global seismic parameters and individual frequencies of the modes. The correlation between the $\epsilon$ parameter and the effective temperature of the stars is used in this process to verify the correct identification of the $\ell$=0 modes.

By combining the seismic parameters with the effective temperature and luminosity provided by the {\it Gaia} DR2 mission, we searched for the best-fit models with the MESA code, which allows us to estimate the fundamental parameters of our targets. However we remind the reader that we did not use any constraints on metallicity for three of the targets. Our work can be summarized as follows:


   \begin{enumerate}


      \item The computation of the prediction of the location of the p modes based on the {\it Gaia} DR2 data agrees with the observations within 1-$\sigma$ for our sample of solar-like stars.

      \item We compared the frequencies of the modes fitted in this work with the ones from \citet{2021ApJ...922...18O} for the five stars in common. We showed that we fitted additional orders at low and high frequency as well as higher degree modes ($\ell$\,=\,2 or 3). For the modes in common, we find a good agreement within 3\,$\sigma$ in the majority of the cases.
      
      \item The seismic modeling performed for these stars points out that four targets are on the main sequence (HD~116832 (B), HD~115427 (D), HD~120746 (E), and HD~119038 (H)). In addition, two stars are close to have exhausted their hydrogen in the core putting them near the TAMS (HD~114558 (C) and HD~117779 (F)). Finally, HD~119026 (G) is a subgiant and  HD~115680 (A), a red giant. Compared to the previously characterized {\it Kepler} solar-like stars, we find that for a given mass, our evolved stars are among the oldest compared to those observed during the \emph{Kepler} main mission. 
      
      \item For the modeling analysis, we used two grids of models with and without treatment of diffusion. By comparing the fundamental parameters ($T_{\rm eff}$, $\log g$, $M$, $R$, and age) obtained with the two different grids, we find that they all agree within the internal uncertainties, except the age. The ages obtained using the models without diffusion are older (Fig.~\ref{age_dif_ndif_plot}) with differences greater than 10\% for most of the stars. Indeed, microscopic diffusion affects the abundance of chemical elements inside stars, yielding to a change in the amount of H in the core. As a consequence ages computed with the models with diffusion are smaller compared to the ones computed without diffusion.
      
      \item The comparison between the radii from the seismic modeling (computed this time without the luminosity constraint from {\it Gaia} this time) and the {\it Gaia} DR2 radii shows that on average they differ by 4\% with a dispersion of 5\%. We find that in general the seismic radii are slightly underestimated, with the largest disagreement for more evolved stars, which is consistent with previous comparisons between {\it Gaia} and asteroseismology. 
      
      \item Using the C6 EVEREST light curves, we also looked for signatures of rotation via spot modulations and/or faculae in the K2 observations. We find two stars with potential rotation periods (HD~120746 (E) and HD~117779 (F)). However, additional observations are needed in order to confirm these modulations as the real rotation periods.
      
      \item On the study of the variation of the p-mode parameters with time, it was not possible to uncover any significant frequency shift due to the high uncertainties on the frequencies of the main-sequence stars. An upper limit of $\sim$0.4$\mu$Hz could be considered during the three years interval between the K2 observations.

   \end{enumerate}

In order to improve the stellar parameters of our sample of stars, additional high-resolution spectroscopic observations would be very useful (in particular to have reliable constraints on the metallicity for the stars without such constraints). In addition, to make an in-depth study of their rotation periods, new observation campaigns are required. This could be achieved with the NASA Transiting Exoplanet Survey Satellite \citep[TESS][]{2015JATIS...1a4003R} mission, currently in operation, and the future PLAnetary Transits and Oscillations of stars \citep[PLATO][]{2014ExA....38..249R} from the European Space Agency.

\begin{acknowledgements}
      This paper includes data collected by the \emph{Kepler} mission. Funding for the \emph{Kepler} mission is provided by the NASA Science Mission directorate. Some of the data presented in this paper were obtained from the Mikulski Archive for Space Telescopes (MAST). STScI is operated by the Association of Universities for Research in Astronomy, Inc., under NASA contract NAS5-26555. 
      LGC acknowledges support from grant FPI-SO from the Spanish Ministry of Economy and Competitiveness (MINECO) (research project SEV-2015-0548-17-2 and predoctoral contract BES-2017-082610). 
      S.M. and L.G.C acknowledge support from the Spanish Ministry of Science and Innovation (MICINN) with the grant no.~PID2019-107061GB-C66 for PLATO.
      S.M. and and D.G.R. acknowledge support from the Spanish Ministry of Science and Innovation (MICINN) with the grant no. PID2019-107187GB-I00. S.M. also acknowledges support from MICINN with the Ram\'on y Cajal fellowship no.~RYC-2015-17697 and through AEI under the Severo Ochoa Centres of Excellence Programme 2020--2023 (CEX2019-000920-S).
      R.~A.~G. and S.~N.~B acknowledge the support from PLATO and GOLF CNES grants. The paper made use of the IAC Supercomputing facility HTCondor (http://research.cs.wisc.edu/htcondor/), partly financed by the MINECO with FEDER funds, code IACA13-3E-2493. The targets analyzed in this work were part of the K2 Guest Observer proposal numbers GO6039 and GO17036 led by Guy Davies and Mikkel Lund.\\
      \textit{Software}: Python \citep{2013A&A...558A..33A}, NumPy \citep{book, Harris:2020ti}, matplotlib \citep{4160265}, astropy \citep{2013A&A...558A..33A}, pandas \citep{reback2020pandas, proc-scipy-2010}, emcee \citep{2013PASP..125..306F}, KADACS \citep{2011MNRAS.414L...6G}, A2Z \citep{2010A&A...511A..46M}, Apollinaire \citep{2022A&A...663A.118B}, lightkurve \citep{2018ascl.soft12013L}, EVEREST \citep{2016AJ....152..100L,2018AJ....156...99L}.
      
\end{acknowledgements}

\bibliographystyle{aa} 
\bibliography{./BIBLIO_sav_v1} 




\begin{appendix}
\onecolumn
\renewcommand\thefigure{\thesection.\arabic{figure}}

\section{Comparison between EVEREST and Lightkurve}\label{appendix:A}

In order to compare EVEREST and Lightkurve calibration pipelines, we study the background and universal pattern parameters from \texttt{apollinaire} for the C6 data, as we have both extracted light curves ($\rm C6_{\rm EV}$ and $\rm C6_{\rm LK}$). The comparison should mainly show differences between the calibration methods as the stellar signal should be roughly the same in both datasets. 

\begin{figure}[h]
    \centering
    \includegraphics[width=0.8\textwidth]{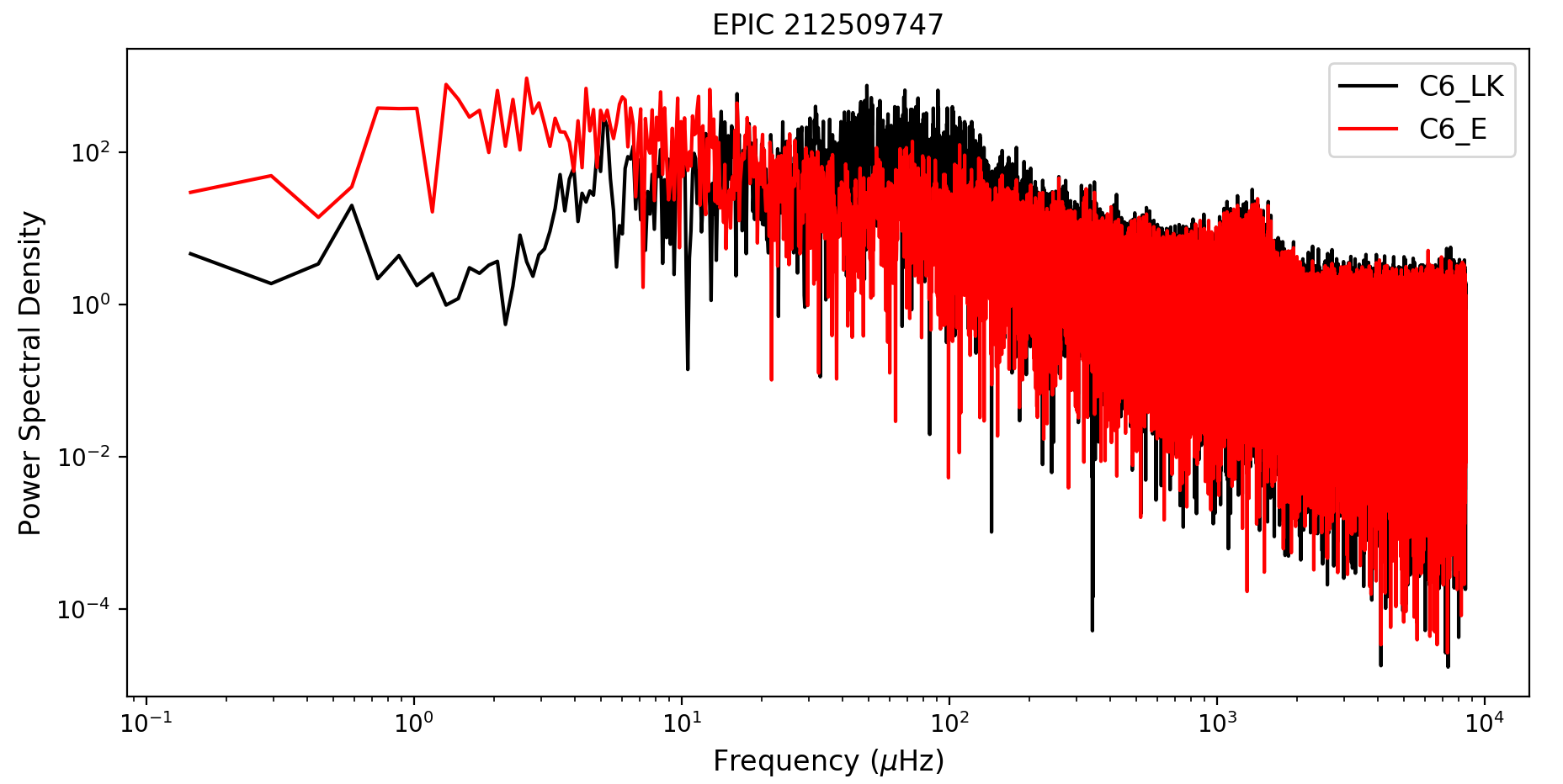}
    \caption{Power spectrum density of the HD~115427 (D) obtained for C6 light curves calibrated with EVEREST (in red) and with the Lightkurve adapted package (in black).}
    \label{EPIC747_psd}
\end{figure}

In Figure~\ref{EPIC747_psd}, we show the comparison of the PSD obtained from both calibrations for HD~115427 (D). The first noticeable result is the difference at low frequency, $\nu$ < 200 $\mu$Hz, related to the filter that is used in Lightkurve as explained in Section~\ref{sec:2}. We remind that the \texttt{apollinaire} analysis we have performed only takes into account the PSD above 50\,$\mu$Hz. 

For a more quantitative analysis, we first compare the background parameters of both Harvey models that we fitted: frequency ($\nu_{\rm c,Hi}$) and amplitude ($A_{Hi}$). We find that for all stars both parameters disagree by more than 1\,$\sigma$.
In the comparison of the noise parameter, a much higher noise is obtained for the light curve from $\rm C6_{\rm LK}$.

For the case of HD~115427 (D), in addition to differences in the background parameters, the frequency of maximum power also differs by more than 1\,$\sigma$. In the PSD, we can see that the p-mode bump is wider with EVEREST compared to Lightkurve. Besides, the fit of the Gaussian envelope is done on top of the background. If the backgrounds are different, it will also impact the fit of the Gaussian function on the p-mode envelope.


For the subgiant, HD~119026 (G), differences larger than 1\,$\sigma$ are also observed in the comparison between the parameters $\nu_{\rm c,H2}$ and $A_{\rm Gauss}$. $\rm W_{\rm env}$ is the width of the Gaussian envelope of the p modes. This larger difference for HD~115427 (D) indicates us that the width of the envelope is greater in $\rm C6_{\rm EV}$ than in $\rm C6_{\rm LK}$.

In general, we found that the signal-to-noise ratio (SNR) is higher for the EVEREST lightcurves.


Finally, we look at the comparison of the parameters obtained from the Universal Pattern fit in \texttt{apollinaire}: $\epsilon$, $\alpha$, $\delta \nu$, $\nu_{\rm max}$ and $W_{\rm env}$. We find no significant differences between the two calibrations, which usually lay below the 3\,$\sigma$ level. 


To conclude, we found that the noise is higher in the light curve calibrated with Lightkurve compared to the EVEREST light curve. Therefore, we decided to use $\rm C6_{\rm EV}$ light curve and combine it with C17$_{\rm LK}$ light curve for the full analysis of the K2 data available for our targets.

\section{Effect of the gap removal on EPIC~212485100}
\label{appendix:gap}

Removing a long gap in the light curve produces a sudden break in the phase of the modes. On one hand, if the lifetime of the modes is shorter than the gaps, the phase of the mode is already broken and there is no effect on the mode's properties. On the other hand, if the modes are long lived, that is longer than the gap, the phases are still coherent and the lifetime of the modes will be artificially reduced to the length of the segments. In other words, the modes will be  
widened. This effect has been thoroughly studied using Monte-Carlo simulations as well as long solar datasets in Section 4 of \cite{2004ESASP.538..265B} and in Section 5 of \cite{2004A&A...423.1051B}. For the stars analyzed in this work, the thinnest modes have widths of around 1 $\mu$Hz, corresponding to lifetimes of $\sim$3.7 days \citep[see for example eq.~24 in][]{2019LRSP...16....4G}. It is then clear that during the $\sim$880 days of the gap the modes have been re-excited hundreds of times and thus the break in the phase that we have imposed by removing the gap has no effect on the limit spectrum of the modes.  

Recently, in the paper by \cite{2022RNAAS...6..202B}, the authors investigated the impact of a gap in photometric data on solar-like oscillations characterization. However, the authors did not correctly interpret the effect of the limit spectrum and the excitation function. In the seismic analysis of solar-like stars with stochastically excited modes, what is important is to characterize the underlying limit spectrum (the Lorentzian function) and not the effect of the excitation that multiplies it. Therefore, it is absolutely normal that the distribution of points around the limit Lorentzian function is different when comparing PSDs of the gap removed series with the ones with a gap. When a Lorentzian fitting is performed on the PSD, the properties of the function are correctly retrieved whatever the exciting function looks like.

\begin{figure}[!htb]
    \centering
    \includegraphics[width=0.8\textwidth]{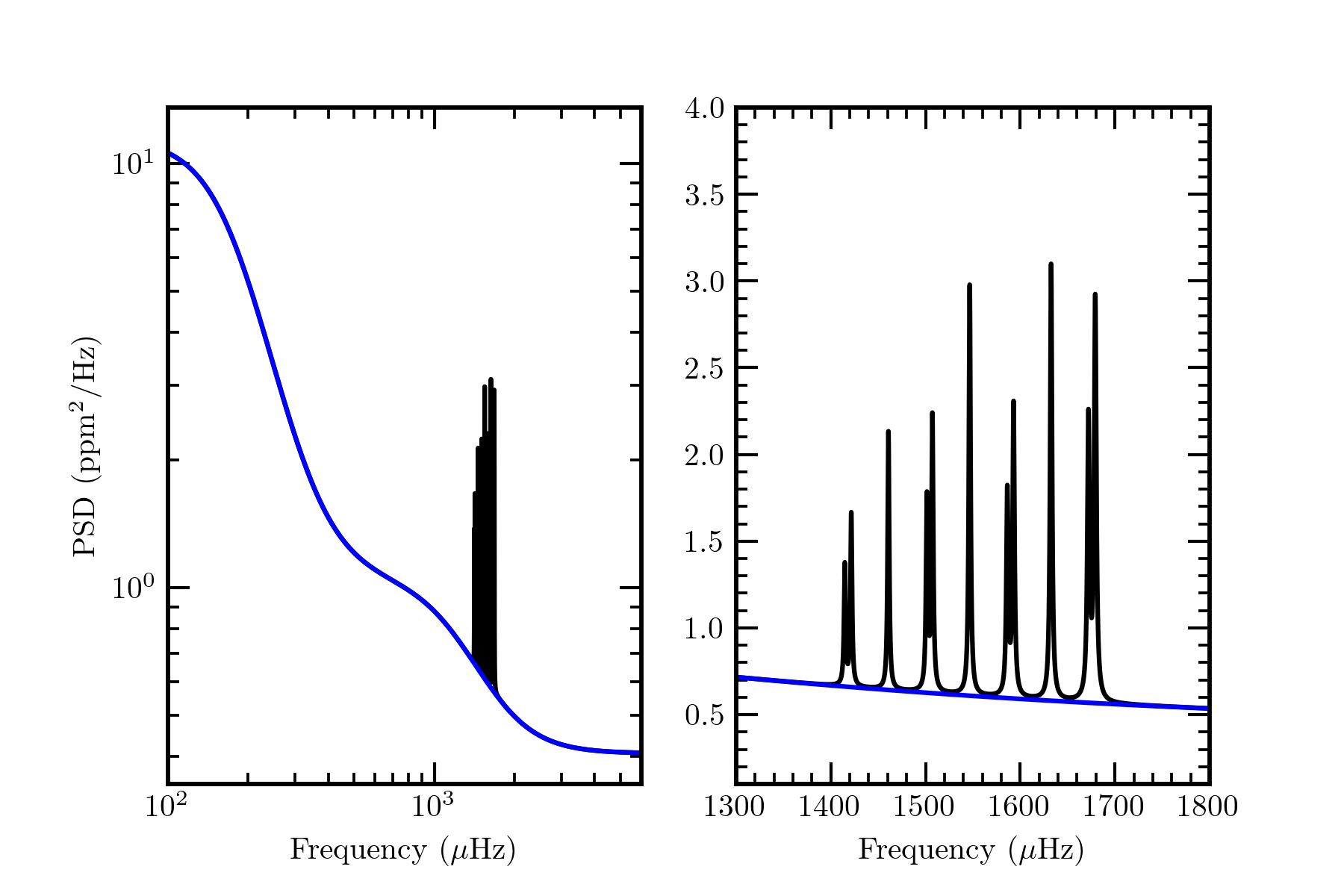}
    \caption{The left panel represents the limit spectrum of the 11 simulated modes based on EPIC~212485100 (black lines) as well as the contribution of the convective background and the photon noise (blue line). The right panel is a zoom around the p modes.}
    \label{synth_spec}
\end{figure}

Although \cite{2004ESASP.538..265B,2004A&A...423.1051B} already proved that it is perfectly correct to remove the gap for short-lived stochastic excited modes and study the resultant time series as a whole, we present here the analysis of simulated data based on one of the K2 targets, EPIC~212485100. We performed a 500 Monte-Carlo simulation of the 11 longest lived modes in EPIC~212485100 (from l=2 at $\nu=$ 1414.76 $\mu$Hz up to l=0 at $\nu=$1679.24 $\mu$Hz) in the  same conditions as in the K2 data analyzed in this paper, that is a total time span of 1029 days consisting of a first series of 79 days (similar to the C6 observations), a gap of 883 days and a final segment of 67 days (as in C17). The properties of the modes and the background to build the limit spectrum were taken from the fit of the C06 EVEREST dataset. In Fig.~\ref{synth_spec}, the limit spectrum used to compute the Monte-Carlo simulation is shown. Once this limit spectrum is calculated we convert the PSD into an amplitude spectrum (AS) and we multiply the real and imaginary parts of the AS by two Gaussian noise distributions. By calculating the inverse Fourier transform provides us with the 1079-day light curve associated with a single noise realization. Then we repeat the process 500 times.

\begin{figure}[!htb]
    \centering
    \includegraphics[width=0.8\textwidth]{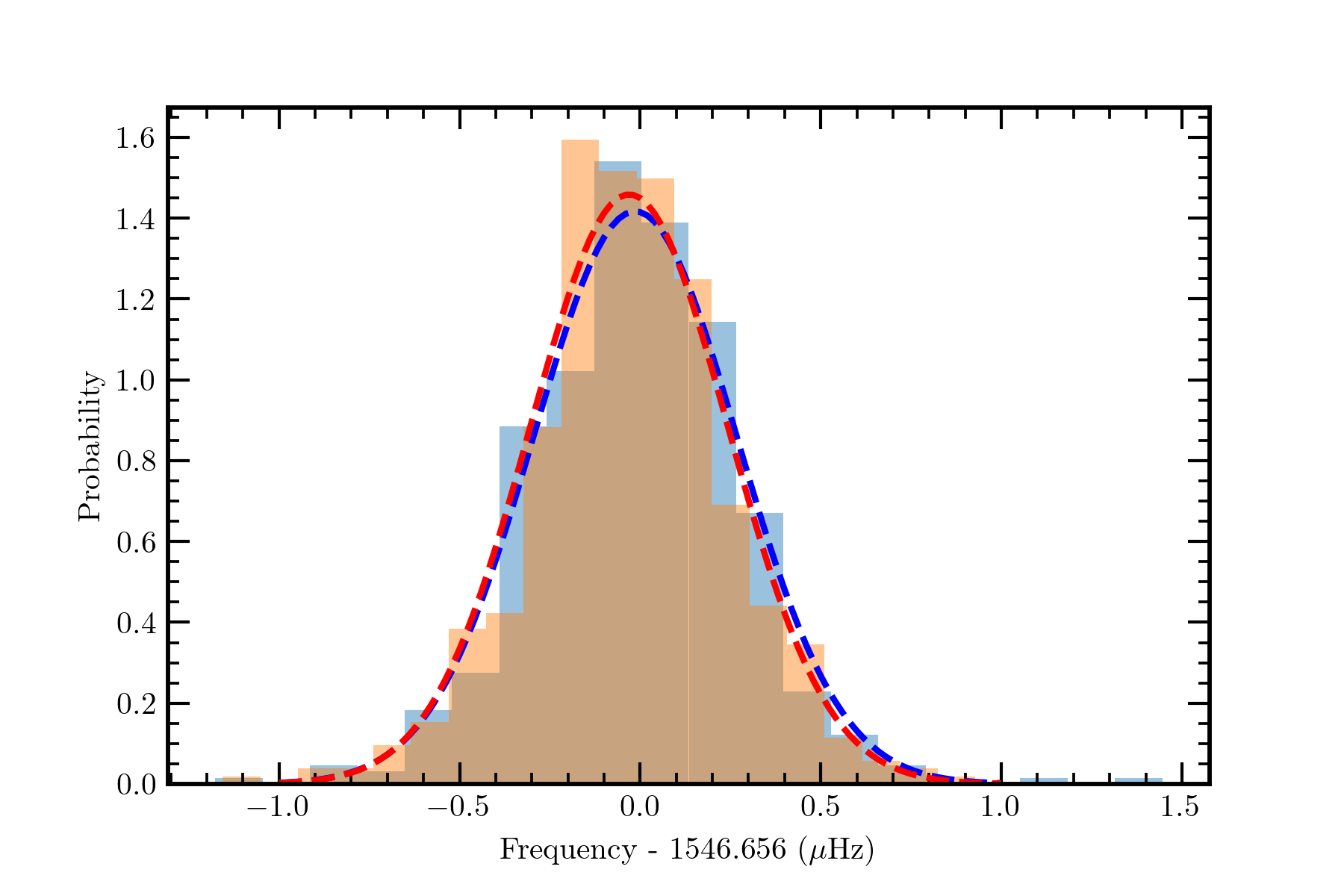}
    \caption{Histogram of the 500 Monte-Carlo simulations of the frequencies of the fitted mode $ell=$1, $n$=16  after subtracting the theoretical value of $\nu=$1546.656 $\mu$Hz. In blue the results of the gap-corrected series. In orange the result of the light curve of the consecutive 146-day series. The dashed lines are the fitted Gaussian distributions as explained in the text.}
    \label{synth_histo}
\end{figure}

In the following, we describe the results obtained when studying the mode $\ell=$1, $n=$16 at $\nu=$1546.656 $\mu$Hz, which is placed at the center of the 11 simulated modes. The analysis of the other modes provides similar conclusions. Two cases have been simulated: a) the analysis of the first continuous 146 days of the simulated time series, that is, there is no forced break of the phase and it will be used as a reference to compare with; b) the analysis of the last 67 days segment concatenated with the first segment after removing the gap for a total of 146 days as done in the analysis of the K2 real data. For both cases the fitting with the \texttt{apollinaire} code is performed. In Fig.~\ref{synth_histo}, the distribution of the fitted frequency (subtracted by the theoretical value of 1546.656 $\mu$Hz) is shown, in orange for the continuous series with the red dashed line representing the fit of a Gaussian distribution to the data. In blue, are represented the results for the gap-removed series with the blue dashed line representing the theoretical Gaussian fit. The mean and the standard deviation of the Gaussian functions are  -0.03 and 0.27, and -0.014 and 0.28 respectively for both a) and b) cases. Thus, the difference of the center of the distributions is zero inside the statistical uncertainties. There is no bias in the center of the distribution of the gap-removed series. Moreover the standard deviation is also the same proving that removing the gap has no effect in the characterization of the limit spectrum while having nearly twice the frequency resolution compared to either the analysis of each campaign separately or the average of both of them.

It is also important to emphasize that averaging the two campaigns is not a good solution as suggested by \cite{2022RNAAS...6..202B}. As already mentioned, by doing so, the resolution will be degraded compared with the analysis of the concatenation of the two campaigns. The second problem would be associated with the change of the statistics when averaging two PSDs. In this case, the statistics of the averaged PSD is not a $\chi^2$ with 2 d.o.f. but a higher degree of freedom. As a consequence, the error bars obtained from the fit need to be corrected as explained in \cite{2003A&A...412..903A}. Unfortunately, the correction factor provided in this last paper has been obtained for a Maximum likelihood minimization while we are using a Bayesian approach. It is out of the scope of this paper to study this procedure in detail when an alternative solution has been proved to be correct. Moreover, as a zero padding is needed in the second campaign to reach the same length of 79 days as the first time series in order to have the same frequency resolution in the PSDs before averaging them, this implies the addition of a correlation between the points. Hence, the resulting statistical distribution of the PSD of the second campaign is already a $\chi^2$ with a higher degree of freedom that will then modify the statistics of the final averaged PSD in a very different way from what was studied by \cite{2003A&A...412..903A}. As a  conclusion, it is not possible to asses the reliability of the inferred error bars in the Bayesian fit when averaging the PSDs of the two K2 campaigns considered here. Therefore, averaging the PSDs is not a valid methodology in our study.

    \newpage

\section{Individual frequencies of the p modes}
\label{appendix:freq}
In this appendix, we provide the tables with the results of the \texttt{apollinaire} fit for each star: the radial order, n, the mode degree, $\ell$, and the frequency of the mode, $\nu_{l,n}$, with the associated error. Mixed modes are those with an order n\,$>$\,100. The column ``flag'' provides the list of modes used to find the best-fit stellar evolution models. A flag equals to 1 indicates that the mode was used, while 0 means that the mode was not used. The last column (``Fitted by Ong+21'')  indicates whether the mode was fitted in \citet{2021ApJ...922...18O}.  Modes that were fitted by the latter have ``Fitted by Ong+21 = 1'' and new modes from our analysis are indicated by ``Fitted by Ong+21 = 0''. 

\begin{table*}[h!]
\label{table:A1} 
\centering 
\caption{Individual frequencies for HD~115680 (A, EPIC~212478598).}
\begin{tabular}{c c c c c c}
\hline\hline 
 l & n & $\nu_{l,n}$ & err $\nu_{l,n}$ & flag & Fitted by Ong+21 \\ 
\hline 
0 &  11 & 403.5692 & 0.0621 & 1 & 0 \\
0 &  12 & 437.7208 & 0.0442 & 1 & 0 \\
0 &  13 & 472.5090 & 0.0465 & 1 & 1 \\
0 &  14 & 507.5396 & 0.0404 & 1 & 1 \\
0 &  15 & 542.4378 & 0.0452 & 1 & 1 \\
0 &  16 & 577.6363 & 0.0778 & 1 & 1 \\
0 &  17 & 612.9136 & 0.1993 & 1 & 1 \\
0 &  18 & 648.3358 & 0.3572 & 1 & 0 \\
1 & 101 & 409.9749 & 0.1364 & 0 & 0 \\
1 &  11 & 421.4803 & 0.0865 & 0 & 1 \\
1 & 102 & 451.9484 & 0.0479 & 0 & 1 \\
1 &  12 & 461.9758 & 0.0676 & 0 & 1 \\
1 &  13 & 493.1587 & 0.0619 & 1 & 1 \\
1 & 104 & 514.6980 & 0.0737 & 0 & 1 \\
1 &  14 & 527.2039 & 0.0556 & 1 & 1 \\
1 & 105 & 549.1423 & 0.0880 & 0 & 1 \\
1 &  15 & 562.5527 & 0.0475 & 1 & 1 \\
1 & 107 & 571.8132 & 0.1153 & 0 & 0 \\
1 &  16 & 599.8420 & 0.0584 & 0 & 1 \\
1 & 109 & 626.5412 & 0.0783 & 0 & 1 \\
1 &  17 & 641.5892 & 0.0863 & 0 & 1 \\
1 & 110 & 666.0361 & 0.1646 & 0 & 1 \\
1 & 111 & 706.4134 & 0.1036 & 0 & 1 \\
2 &  11 & 435.1010 & 0.0557 & 1 & 0 \\
2 &  12 & 468.9194 & 0.1013 & 1 & 0 \\
2 &  13 & 503.5965 & 0.0515 & 1 & 0 \\
2 &  14 & 538.8650 & 0.0429 & 1 & 0 \\
2 &  15 & 574.9569 & 0.0838 & 1 & 1 \\
3 &  12 & 482.8612 & 0.2355 & 1 & 1 \\
3 &  13 & 516.3816 & 0.2448 & 1 & 1 \\
3 &  14 & 551.7128 & 0.0862 & 1 & 1 \\
3 &  15 & 587.0085 & 0.0421 & 1 & 0 \\
\hline 
\end{tabular}
\end{table*}
\FloatBarrier

\begin{table*}[h!]
\caption{Individual frequencies for HD~116832 (B, EPIC~212485100).}
\label{table:A2} 
\centering 
\begin{tabular}{c c c c c c}  
\hline\hline 
l & n & $\nu_{l,n}$ & err $\nu_{l,n}$ & flag & Fitted by Ong+21 \\ 
\hline 
0 & 14 & 1334.5835 & 0.3626 & 1 & 0 \\  
0 & 15 & 1421.5567 & 0.3564 & 1 & 0 \\
0 & 16 & 1507.2339 & 0.3841 & 1 & 1 \\
0 & 17 & 1592.9731 & 0.4666 & 1 & 1 \\
0 & 18 & 1679.2417 & 0.2549 & 1 & 1 \\
0 & 19 & 1766.6436 & 0.3677 & 1 & 1 \\
0 & 20 & 1854.7448 & 0.2653 & 1 & 1 \\
0 & 21 & 1942.1498 & 0.3718 & 1 & 0 \\
0 & 22 & 2030.5946 & 0.5733 & 1 & 1 \\
0 & 23 & 2116.6611 & 0.4233 & 1 & 1 \\
0 & 24 & 2204.8282 & 0.7518 & 1 & 1 \\
1 & 13 & 1287.2890 & 0.9119 & 1 & 0 \\
1 & 14 & 1373.9909 & 0.4031 & 1 & 1 \\
1 & 15 & 1460.8275 & 0.2849 & 1 & 1 \\
1 & 16 & 1546.6561 & 0.2057 & 1 & 1 \\
1 & 17 & 1632.6167 & 0.2300 & 1 & 1 \\
1 & 18 & 1719.8929 & 0.2519 & 1 & 1 \\
1 & 19 & 1806.8575 & 0.2240 & 1 & 1 \\
1 & 20 & 1895.6261 & 0.2508 & 1 & 1 \\
1 & 21 & 1982.5715 & 0.3081 & 1 & 1 \\
1 & 22 & 2070.4191 & 0.4429 & 1 & 1 \\
1 & 23 & 2159.9056 & 0.6844 & 1 & 1 \\
2 & 13 & 1327.8703 & 0.5227 & 1 & 0 \\
2 & 14 & 1414.7684 & 0.8608 & 1 & 0 \\
2 & 15 & 1501.4548 & 1.6710 & 1 & 1 \\
2 & 16 & 1586.3712 & 0.4091 & 1 & 0 \\
2 & 17 & 1672.0806 & 0.3988 & 1 & 1 \\
2 & 18 & 1759.8920 & 0.5635 & 1 & 0 \\
2 & 19 & 1848.9506 & 0.4816 & 1 & 0 \\
2 & 20 & 1936.8328 & 0.6702 & 1 & 0 \\
2 & 21 & 2023.2062 & 0.7102 & 1 & 0 \\
2 & 22 & 2112.0564 & 1.4168 & 1 & 0 \\
2 & 23 & 2200.6703 & 1.6832 & 1 & 0 \\
\hline 
\end{tabular}
\end{table*}
\FloatBarrier

\begin{table*}[h!]
\caption{Individual frequencies for HD~114558 (C, EPIC~212487676).}
\label{table:A3} 
\centering 
\begin{tabular}{c c c c c c}  
\hline\hline 
l & n & $\nu_{l,n}$ & err $\nu_{l,n}$ & flag & Fitted by Ong+21 \\ 
\hline 
0 & 12 &  942.5367 & 0.6219 & 1 & 0 \\
0 & 13 & 1018.1983 & 0.5245 & 1 & 0 \\
0 & 14 & 1094.2343 & 0.5100 & 1 & 1 \\
0 & 15 & 1168.3366 & 0.2013 & 1 & 1 \\
0 & 16 & 1243.6447 & 0.2187 & 1 & 1 \\
0 & 17 & 1318.6450 & 0.2085 & 1 & 1 \\
0 & 18 & 1395.3834 & 0.1349 & 1 & 1 \\
0 & 19 & 1471.6646 & 0.2849 & 1 & 1 \\
0 & 20 & 1547.9279 & 0.2528 & 1 & 1 \\
0 & 21 & 1624.2704 & 0.4036 & 1 & 0 \\
0 & 22 & 1702.1244 & 0.9191 & 1 & 0 \\
0 & 23 & 1777.9899 & 0.8971 & 1 & 0 \\
0 & 24 & 1856.0356 & 1.1906 & 1 & 0 \\
1 & 12 &  975.0427 & 0.2682 & 1 & 0 \\
1 & 13 & 1050.7885 & 0.2960 & 1 & 1 \\
1 & 14 & 1126.0174 & 0.5649 & 1 & 1 \\
1 & 15 & 1200.1125 & 0.1111 & 1 & 1 \\
1 & 16 & 1274.8267 & 0.2061 & 1 & 1 \\
1 & 17 & 1351.3930 & 0.2116 & 1 & 1 \\
1 & 18 & 1427.9950 & 0.1892 & 1 & 1 \\
1 & 19 & 1504.5355 & 0.1575 & 1 & 1 \\
1 & 20 & 1580.3156 & 0.2851 & 1 & 1 \\
1 & 21 & 1657.1508 & 0.3781 & 1 & 1 \\
1 & 22 & 1734.7114 & 0.4429 & 1 & 0 \\
1 & 23 & 1811.0685 & 0.5329 & 1 & 0 \\
2 & 11 &  936.1945 & 0.2205 & 1 & 0 \\
2 & 12 & 1012.3049 & 0.5099 & 1 & 0 \\
2 & 13 & 1086.2908 & 0.4557 & 1 & 0 \\
2 & 14 & 1162.9756 & 0.2076 & 1 & 0 \\
2 & 15 & 1237.9923 & 0.3211 & 1 & 0 \\
2 & 16 & 1313.6245 & 0.2955 & 1 & 1 \\
2 & 17 & 1390.0330 & 0.3471 & 1 & 0 \\
2 & 18 & 1466.3261 & 0.2634 & 1 & 1 \\
2 & 19 & 1542.3539 & 0.3203 & 1 & 1 \\
2 & 20 & 1619.1551 & 1.5093 & 1 & 1 \\
2 & 21 & 1696.5240 & 0.9382 & 1 & 0 \\
2 & 22 & 1774.0274 & 1.1987 & 1 & 0 \\
3 & 16 & 1343.0168 & 0.5645 & 1 & 0 \\
3 & 17 & 1420.2915 & 1.1555 & 1 & 0 \\
3 & 18 & 1496.1119 & 0.4205 & 1 & 0 \\
3 & 19 & 1573.0425 & 0.5741 & 1 & 0 \\
\hline 
\end{tabular}
\end{table*}

\begin{table*}[h!]
\caption{Individual frequencies for HD~115427 (D, EPIC~212509747). This star has not been analyzed in \citet{2021ApJ...922...18O}}.
\label{table:A4} 
\centering 
\begin{tabular}{c c c c c}  
\hline\hline 
l & n & $\nu_{l,n}$ & err $\nu_{l,n}$ & flag \\ 
 \hline 
0 & 12 &  866.7764 & 0.5145 & 1 \\
0 & 13 &  931.9530 & 0.4923 & 1 \\
0 & 14 &  997.2269 & 0.7600 & 1 \\
0 & 15 & 1062.8577 & 0.5376 & 1 \\
0 & 16 & 1128.5956 & 0.5499 & 1 \\
0 & 17 & 1193.5334 & 0.3463 & 1 \\
0 & 18 & 1258.3010 & 0.5019 & 1 \\
0 & 19 & 1325.4606 & 0.3886 & 1 \\
0 & 20 & 1392.2920 & 0.4097 & 1 \\
0 & 21 & 1459.1665 & 0.4561 & 1 \\
0 & 22 & 1526.2106 & 0.9897 & 1 \\
0 & 23 & 1593.4766 & 0.7833 & 1 \\
0 & 24 & 1660.6746 & 0.8960 & 1 \\
0 & 25 & 1727.0262 & 1.0823 & 1 \\
1 & 10 &  768.1457 & 0.5605 & 1 \\
1 & 11 &  831.8790 & 1.3771 & 1 \\
1 & 12 &  895.3196 & 0.5957 & 1 \\
1 & 13 &  959.5635 & 0.4845 & 1 \\
1 & 14 & 1027.2687 & 0.4725 & 1 \\
1 & 15 & 1091.7071 & 0.3206 & 1 \\
1 & 16 & 1157.3273 & 0.4380 & 1 \\
1 & 17 & 1222.0243 & 0.3398 & 1 \\
1 & 18 & 1287.7555 & 0.3348 & 1 \\
1 & 19 & 1354.8078 & 0.3344 & 1 \\
1 & 20 & 1421.5371 & 0.2804 & 1 \\
1 & 21 & 1489.6142 & 0.4377 & 1 \\
1 & 22 & 1556.0386 & 0.4800 & 1 \\
1 & 23 & 1624.2185 & 0.7436 & 1 \\
1 & 24 & 1690.8360 & 0.8274 & 1 \\
1 & 25 & 1759.5805 & 0.7902 & 1 \\
2 & 13 &  992.8760 & 1.8783 & 1 \\
2 & 14 & 1055.9054 & 1.0255 & 1 \\
2 & 15 & 1124.4532 & 0.7168 & 1 \\
2 & 16 & 1189.2143 & 0.5144 & 1 \\
2 & 17 & 1253.4640 & 0.6150 & 1 \\
2 & 18 & 1319.8505 & 0.9247 & 1 \\
2 & 19 & 1385.9061 & 0.5681 & 1 \\
2 & 20 & 1453.8031 & 0.7326 & 1 \\
2 & 21 & 1521.4784 & 0.7379 & 1 \\
2 & 22 & 1590.1806 & 1.3654 & 1 \\
2 & 23 & 1655.4111 & 1.1260 & 1 \\
2 & 24 & 1724.3426 & 1.4377 & 1 \\
\hline 
\end{tabular}
\end{table*}

\begin{table*}[h]
\caption{Individual frequencies for HD~120746 (E, EPIC~212516207).}
\label{table:A5} 
\centering 
\begin{tabular}{c c c c c c}  
\hline\hline 
l & n & $\nu_{l,n}$ & err $\nu_{l,n}$ & flag & Fitted by Ong+21 \\ 
\hline 
0 & 13 &  894.4540 & 0.8417 & 1 & 0 \\
0 & 14 &  961.5840 & 0.9020 & 1 & 0 \\
0 & 15 & 1029.1293 & 0.3737 & 1 & 1 \\
0 & 16 & 1096.2299 & 0.7332 & 1 & 1 \\
0 & 17 & 1163.8525 & 0.8238 & 1 & 1 \\
0 & 18 & 1230.0923 & 0.8340 & 1 & 0 \\
0 & 19 & 1297.1949 & 0.6113 & 1 & 0 \\
0 & 20 & 1365.5551 & 0.5667 & 1 & 1 \\
0 & 21 & 1433.7355 & 0.2997 & 1 & 1 \\
0 & 22 & 1502.4723 & 0.5499 & 1 & 1 \\
0 & 23 & 1571.5240 & 0.6585 & 1 & 0 \\
0 & 24 & 1639.7364 & 1.1193 & 1 & 0 \\
1 & 13 &  924.6516 & 0.7179 & 1 & 0 \\
1 & 14 &  991.6974 & 0.3746 & 1 & 1 \\
1 & 15 & 1059.7676 & 0.2687 & 1 & 1 \\
1 & 16 & 1127.2197 & 0.3292 & 1 & 1 \\
1 & 17 & 1195.0664 & 0.2744 & 1 & 1 \\
1 & 18 & 1261.5085 & 0.3010 & 1 & 1 \\
1 & 19 & 1328.6056 & 0.3165 & 1 & 0 \\
1 & 20 & 1396.9194 & 0.3024 & 1 & 1 \\
1 & 21 & 1465.4696 & 0.2574 & 1 & 1 \\
1 & 22 & 1535.0859 & 0.4852 & 1 & 1 \\
1 & 23 & 1604.6183 & 0.4628 & 1 & 1 \\
1 & 24 & 1673.4224 & 1.1748 & 1 & 0 \\
2 & 15 & 1091.1862 & 3.0825 & 1 & 0 \\
2 & 16 & 1159.7319 & 0.5426 & 1 & 0 \\
2 & 17 & 1226.6978 & 0.7532 & 1 & 1 \\
2 & 18 & 1293.3876 & 0.6548 & 1 & 0 \\
2 & 19 & 1359.9172 & 0.5876 & 1 & 1 \\
2 & 20 & 1428.5323 & 0.4567 & 1 & 1 \\
2 & 21 & 1498.9018 & 0.9549 & 1 & 1 \\
2 & 22 & 1566.5186 & 1.2731 & 1 & 0 \\
2 & 23 & 1636.2226 & 1.4819 & 1 & 1 \\
\hline 
\end{tabular}
\end{table*}

\begin{table*}[h]
\caption{Individual frequencies for HD~117779 (F, EPIC~212617037). This star has not been analyzed in \citet{2021ApJ...922...18O}}.
\label{table:A6} 
\centering 
\begin{tabular}{c c c c c}  
\hline\hline 
l & n & $\nu_{l,n}$ & err $\nu_{l,n}$ & flag \\ 
\hline 
0 &  11  &  630.5615 & 0.7251 & 1 \\
0 &  12  &  683.2279 & 2.1116 & 1 \\
0 &  13  &  730.7257 & 1.7102 & 1 \\
0 &  14  &  782.5142 & 0.5218 & 1 \\
0 &  15  &  833.4088 & 0.9787 & 1 \\
0 &  16  &  884.0650 & 0.7422 & 1 \\
0 &  17  &  936.6216 & 0.5554 & 1 \\
0 &  18  &  986.6952 & 0.7000 & 1 \\
0 &  19  & 1037.1138 & 1.1748 & 1 \\
0 &  20  & 1088.3932 & 0.9838 & 1 \\
0 &  21  & 1139.7125 & 1.1111 & 1 \\
1 &  11  &  658.7815 & 1.1924 & 1 \\
1 &  12  &  708.3552 & 1.0577 & 0 \\
1 & 102  &  746.6192 & 0.5719 & 0 \\
1 &  13  &  758.6845 & 0.8121 & 0 \\
1 &  14  &  806.2541 & 0.8284 & 1 \\
1 &  15  &  857.2875 & 0.9968 & 1 \\
1 &  16  &  909.1936 & 0.8449 & 1 \\
1 &  17  &  960.4339 & 0.5339 & 1 \\
1 &  18  & 1011.3250 & 0.7134 & 1 \\
1 &  19  & 1061.6697 & 0.9620 & 1 \\
1 &  20  & 1110.6618 & 0.6726 & 1 \\
1 & 110  & 1116.8517 & 0.5549 & 0 \\
2 &  10  &  627.4008 & 1.4291 & 1 \\
2 &  11  &  677.7536 & 1.2396 & 1 \\
2 &  12  &  727.4926 & 0.9729 & 1 \\
2 &  13  &  776.1867 & 0.5906 & 1 \\
2 &  14  &  827.9960 & 0.8855 & 1 \\
2 &  15  &  879.1498 & 1.0204 & 1 \\
2 &  16  &  930.4306 & 0.6010 & 1 \\
2 &  17  &  980.9333 & 0.9412 & 1 \\
2 &  18  & 1033.5436 & 2.0607 & 1 \\
2 &  19  & 1084.7998 & 0.8394 & 1 \\
2 &  20  & 1136.3473 & 1.3384 & 1 \\
\hline 
\end{tabular}
\end{table*}

\begin{table*}[h!]
\caption{Individual frequencies for HD~119026 (G, EPIC~212683142).}
\label{table:A7} 
\centering 
\begin{tabular}{c c c c c c}  
\hline\hline 
l & n & $\nu_{l,n}$ & err $\nu_{l,n}$ & flag & Fitted by Ong+21 \\ 
\hline 
0 &  14  &  651.1179 & 0.6382 & 1 & 0 \\
0 &  15  &  694.7473 & 0.3656 & 1 & 1 \\
0 &  16  &  739.5662 & 0.1990 & 1 & 1 \\
0 &  17  &  785.8174 & 0.2145 & 1 & 1 \\
0 &  18  &  831.2739 & 0.3137 & 1 & 1 \\
0 &  19  &  876.0611 & 0.5493 & 1 & 1 \\
0 &  20  &  923.0355 & 0.4384 & 1 & 0 \\
0 &  21  &  968.6024 & 0.6502 & 1 & 0 \\
0 &  22  & 1016.7130 & 0.5687 & 1 & 0 \\
1 & 103  &  627.9086 & 0.2609 & 0 & 1 \\
1 & 104  &  660.0847 & 0.2216 & 0 & 1 \\
1 & 105  &  685.6036 & 0.2290 & 0 & 1 \\
1 & 106  &  721.5171 & 0.1415 & 1 & 1 \\
1 & 107  &  763.8965 & 0.1291 & 1 & 1 \\
1 & 108  &  807.2637 & 0.1395 & 1 & 1 \\
1 & 109  &  845.7840 & 0.1140 & 0 & 1 \\
1 & 110  &  866.7291 & 0.1795 & 0 & 1 \\
1 & 111  &  902.7550 & 0.2274 & 1 & 1 \\
1 & 112  &  946.6960 & 0.4468 & 1 & 0 \\
1 & 113  &  992.1957 & 0.3772 & 1 & 0 \\
1 & 114  & 1038.9491 & 0.5183 & 0 & 0 \\
2 &  13  &  646.7954 & 1.0958 & 1 & 1 \\
2 &  14  &  690.5915 & 0.4515 & 1 & 1 \\
2 &  15  &  735.7962 & 0.2013 & 1 & 1 \\
2 &  16  &  781.8452 & 0.2900 & 1 & 1 \\
2 &  17  &  828.1601 & 0.4329 & 1 & 1 \\
2 &  18  &  873.1440 & 0.6604 & 1 & 0 \\
2 &  19  &  919.8459 & 0.6933 & 1 & 0 \\
2 &  20  &  965.2275 & 0.5370 & 1 & 0 \\
2 &  21  & 1013.9058 & 0.4671 & 1 & 0 \\
\hline 
\end{tabular}
\end{table*}

\begin{table*}[h!]
\caption{Individual frequencies for HD~119038 (H, EPIC~212772187). This star has not been analyzed in \citet{2021ApJ...922...18O}}.
\label{table:A8} 
\centering 
\begin{tabular}{c c c c c}  
\hline\hline 
l & n & $\nu_{l,n}$ & err $\nu_{l,n}$ & flag \\ 
\hline 
0 & 13 & 1251.8965 & 0.2105 & 1 \\
0 & 14 & 1339.1247 & 1.2815 & 1 \\
0 & 15 & 1426.0718 & 0.8138 & 1 \\
0 & 16 & 1514.1426 & 0.8206 & 1 \\
0 & 17 & 1600.0167 & 0.6234 & 1 \\
0 & 18 & 1688.0900 & 0.7760 & 1 \\
0 & 19 & 1774.5851 & 0.6094 & 1 \\
0 & 20 & 1862.9310 & 0.6557 & 1 \\
0 & 21 & 1952.0329 & 0.8011 & 1 \\
0 & 22 & 2040.4140 & 0.7128 & 1 \\
0 & 23 & 2128.5713 & 1.5657 & 1 \\
0 & 24 & 2217.9405 & 1.1554 & 1 \\
0 & 25 & 2307.8925 & 2.1936 & 1 \\
1 & 11 & 1119.1138 & 1.2100 & 1 \\
1 & 12 & 1202.3374 & 0.5270 & 1 \\
1 & 13 & 1291.4101 & 0.1977 & 1 \\
1 & 14 & 1377.2706 & 1.0819 & 1 \\
1 & 15 & 1466.5041 & 1.2038 & 1 \\
1 & 16 & 1552.9550 & 0.9598 & 1 \\
1 & 17 & 1640.5046 & 0.7578 & 1 \\
1 & 18 & 1728.0558 & 0.5053 & 1 \\
1 & 19 & 1815.0500 & 0.5280 & 1 \\
1 & 20 & 1904.2386 & 0.5765 & 1 \\
1 & 21 & 1994.3726 & 0.5388 & 1 \\
1 & 22 & 2083.9392 & 0.6946 & 1 \\
1 & 23 & 2172.3800 & 1.0360 & 1 \\
1 & 24 & 2262.0431 & 0.9329 & 1 \\
1 & 25 & 2353.3455 & 2.5213 & 1 \\
2 & 12 & 1245.8665 & 0.4730 & 1 \\
2 & 13 & 1331.8571 & 2.1334 & 1 \\
2 & 14 & 1419.6329 & 0.8681 & 1 \\
2 & 15 & 1504.1854 & 0.8245 & 1 \\
2 & 16 & 1592.8977 & 1.0461 & 1 \\
2 & 17 & 1681.6108 & 1.6694 & 1 \\
2 & 18 & 1769.0141 & 1.3032 & 1 \\
2 & 19 & 1857.0122 & 1.2545 & 1 \\
2 & 20 & 1946.9844 & 1.0845 & 1 \\
2 & 21 & 2032.9598 & 1.7167 & 1 \\
2 & 22 & 2122.5536 & 1.6166 & 1 \\
2 & 23 & 2212.8495 & 1.4043 & 1 \\
2 & 24 & 2301.1681 & 1.8098 & 1 \\
\hline 
\end{tabular}
\end{table*}
\FloatBarrier

\newpage

\section{Echelle diagrams and models}\label{appendix:ED}
In this appendix we show the echelle diagrams of the eight stars analyzed in this work, with the modes fitted by \texttt{apollinaire} (red circles) and the frequencies of the best-fit model (orange diamonds). For the model frequencies, only the most reliable modes used in the model fitting are represented as explained in Section\ref{sec:res_stellparam}. 

\begin{figure}[htb!]
    \includegraphics[width=0.5\linewidth]{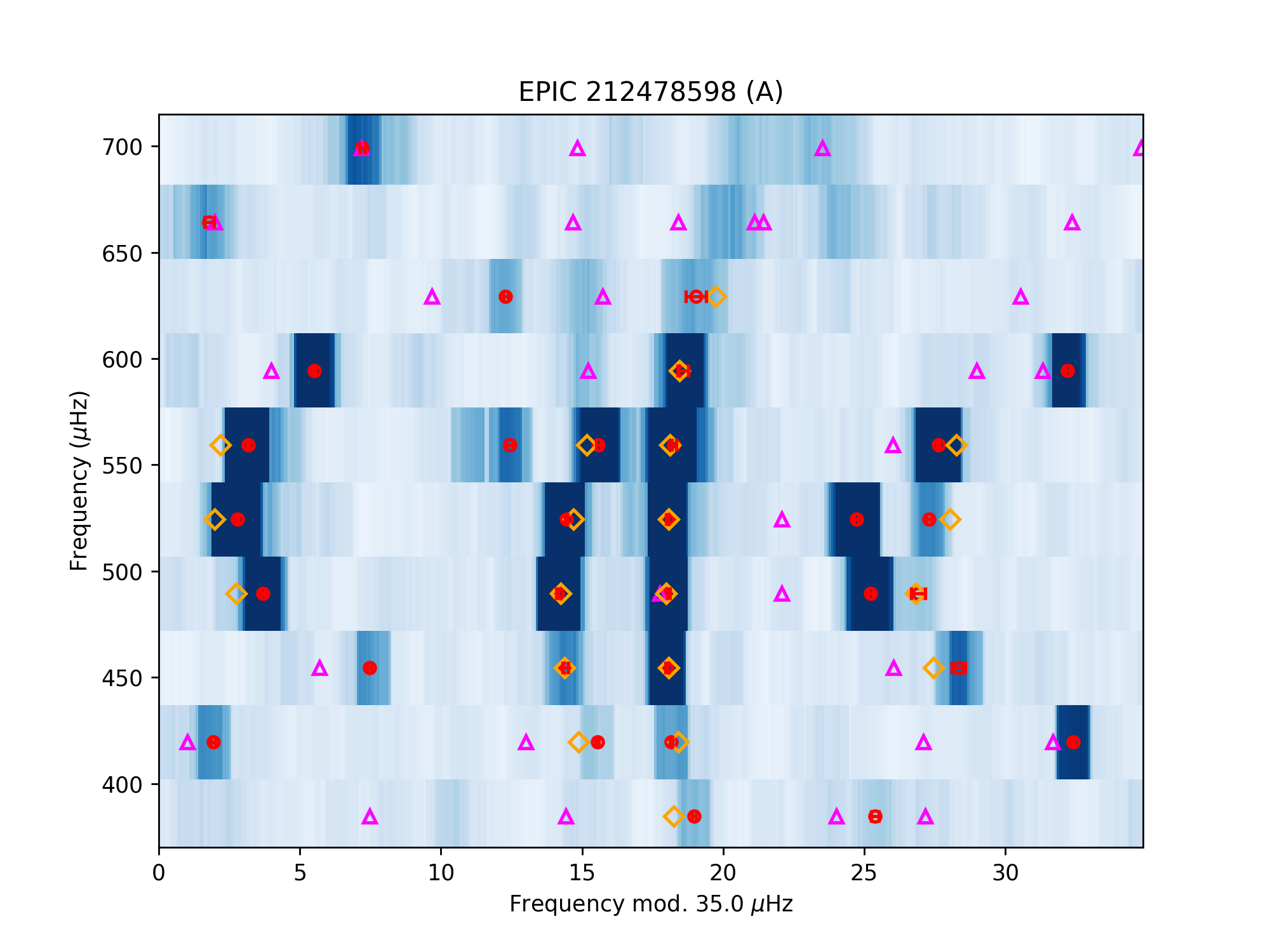}
    \includegraphics[width=0.5\linewidth]{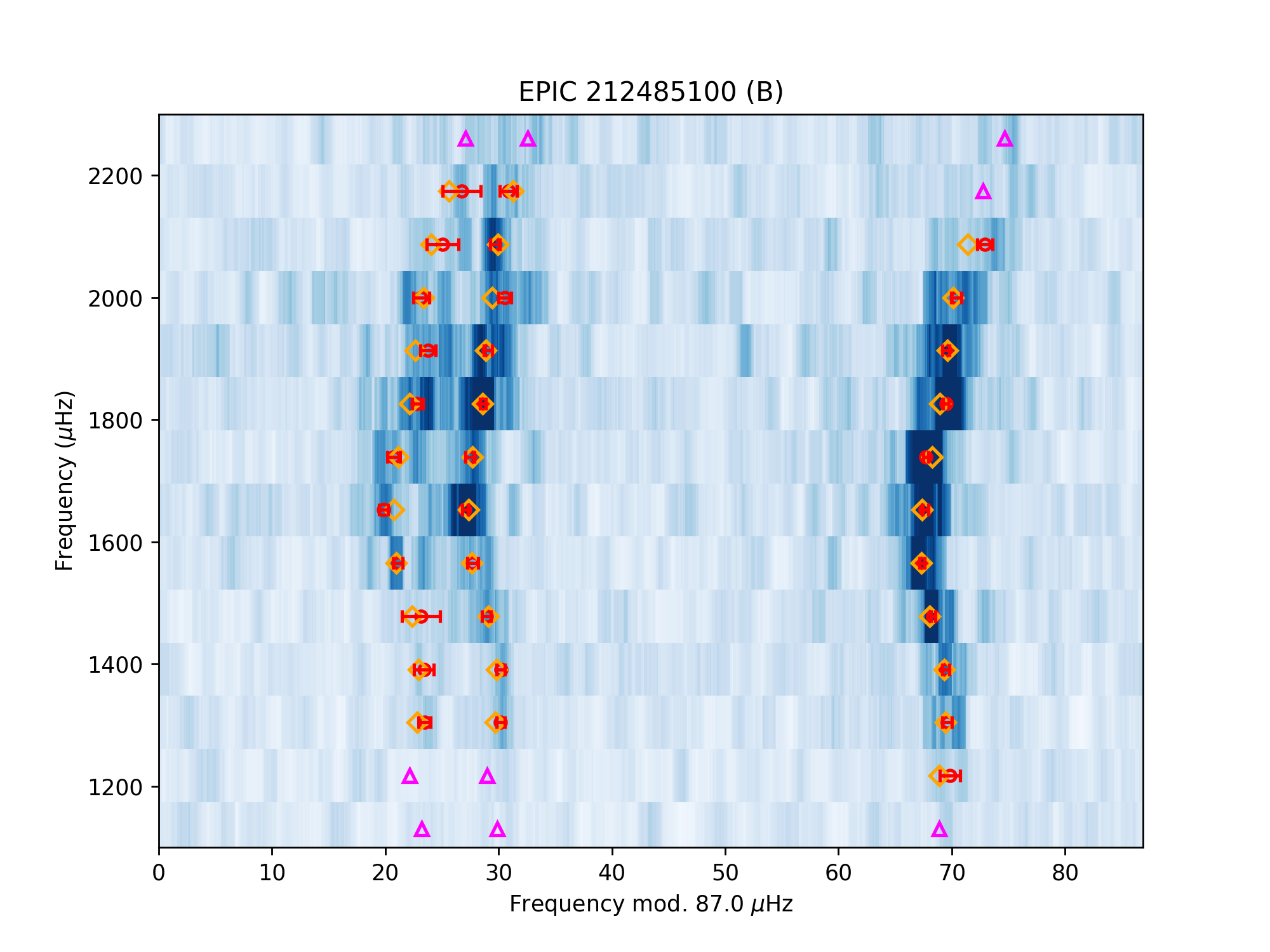}\\
    \includegraphics[width=0.5\linewidth]{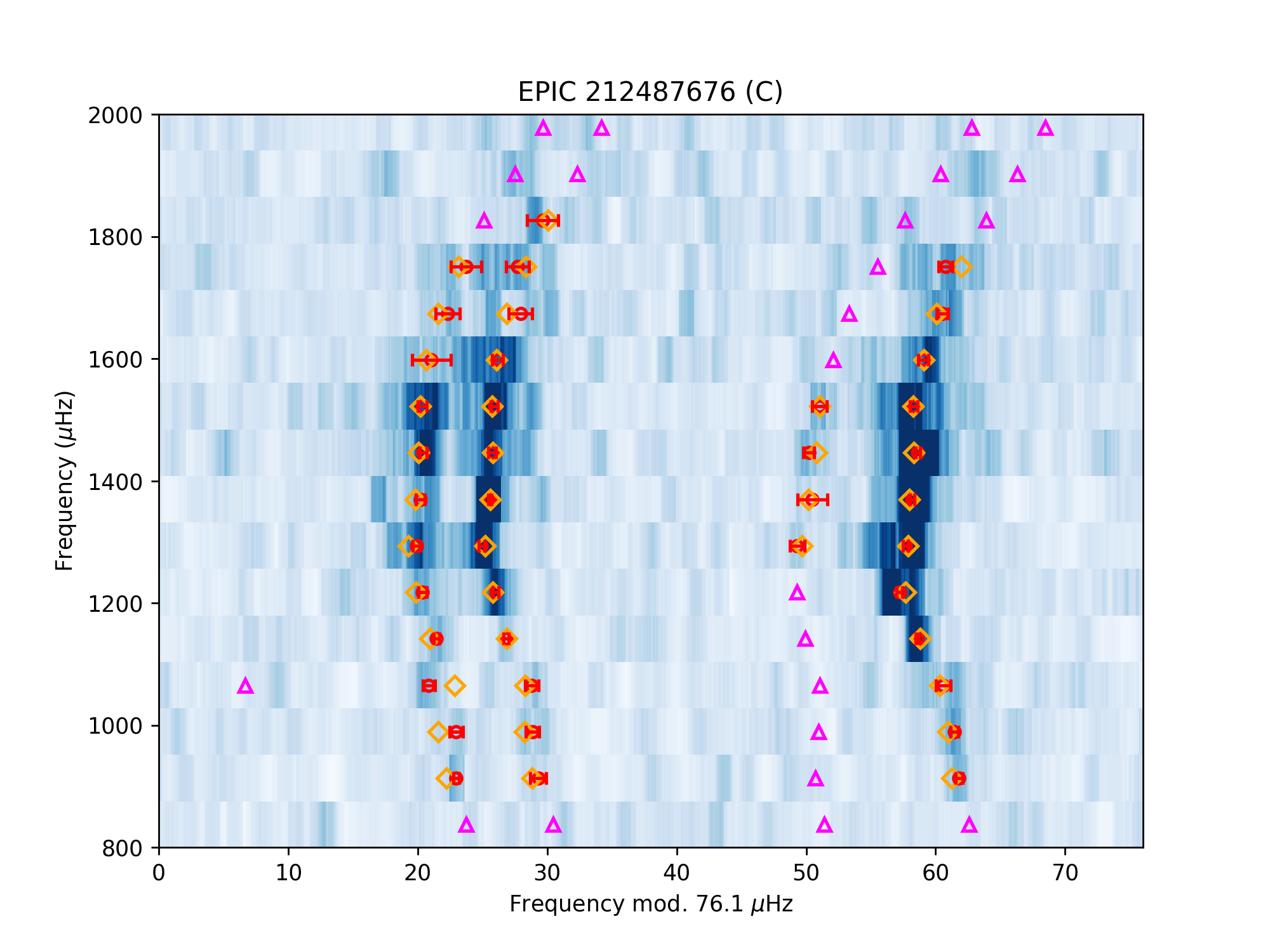}
    \includegraphics[width=0.5\linewidth]{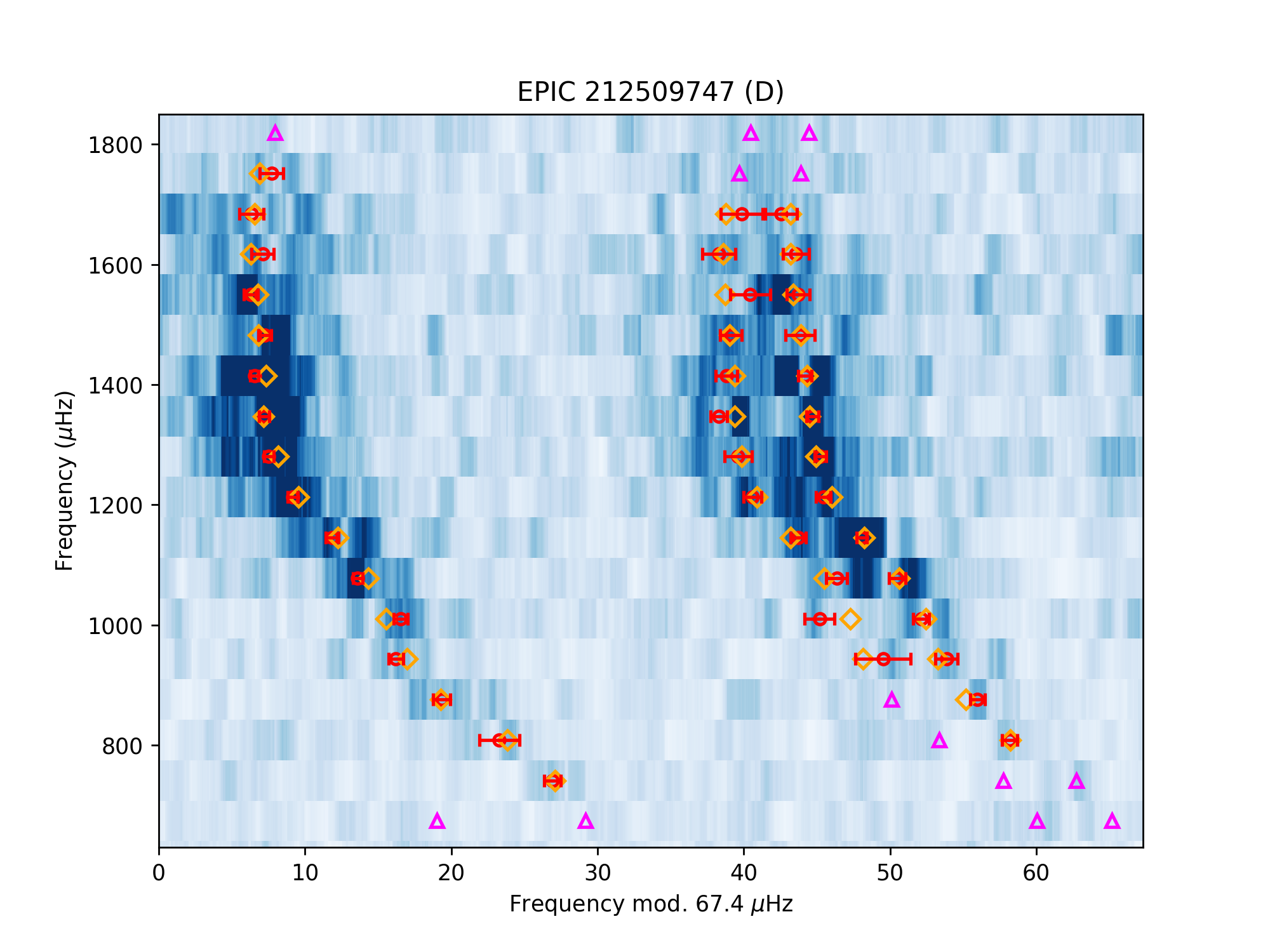}\\
    \caption{Echelle diagram of HD~115680 (A, EPIC~212478598) (top left panel), HD~116832 (B, EPIC~212485100) (top right panel), HD~114558 (C, EPIC~212487676) (bottom left panel) and HD~115427 (D, EPIC~212509747) (bottom right panel). The modes fitted by \texttt{apollinaire} are represented by red circles, while the orange diamonds correspond to the frequencies of the best-fit model after adding the surface corrections for the modes used as input. The magenta triangles represent the theoretical modes that were not part of the model fitting procedure. Also, for the theoretical $\ell$= 2 and 3 modes of HD 115680 (A, EPIC~212478598) star, only modes with a predominant p mode character are shown.} 
    \label{fig:echelle_ABCD}
\end{figure}

\begin{figure}[htb!]
    \includegraphics[width=0.5\linewidth]{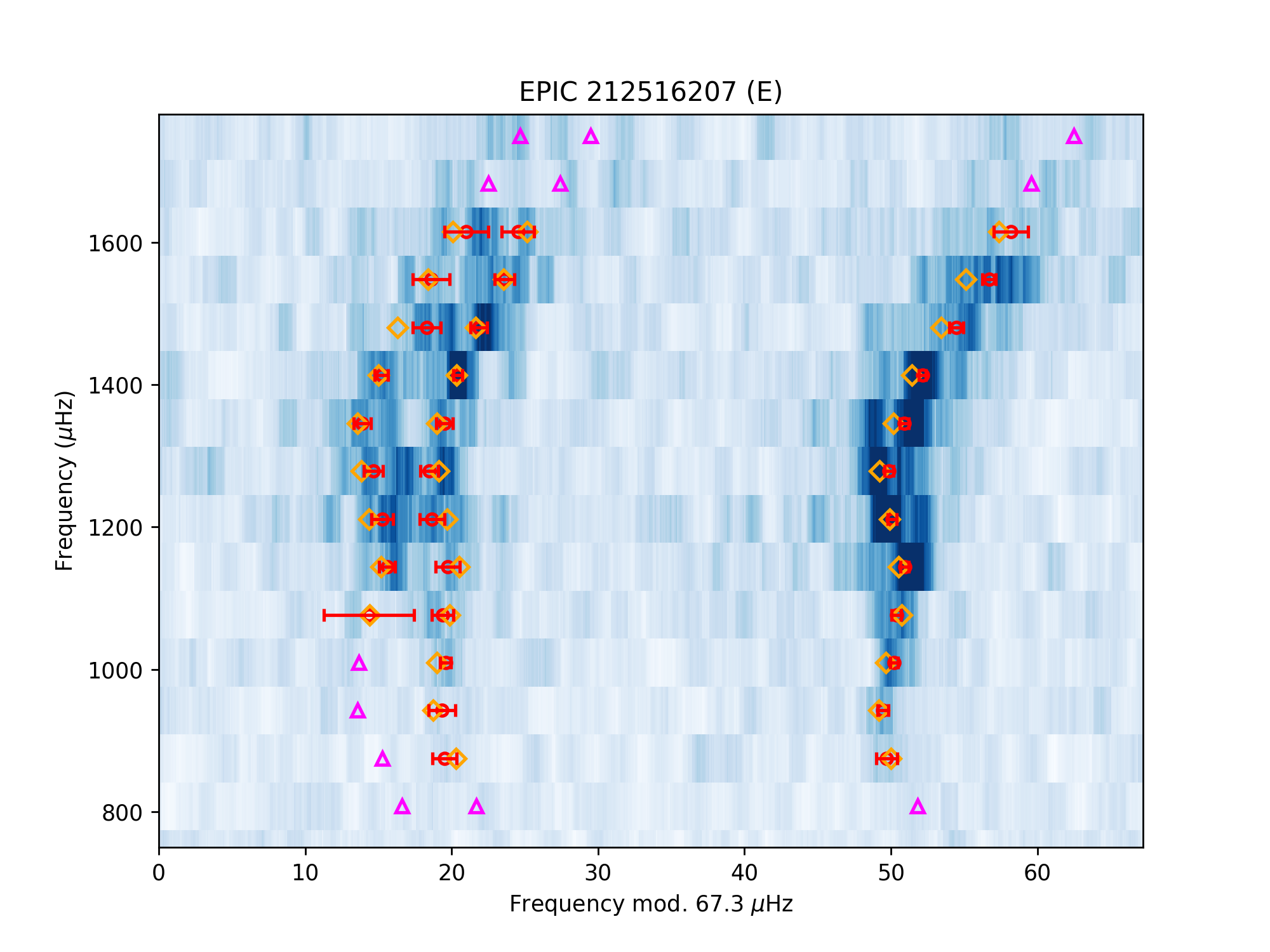}
    \includegraphics[width=0.5\linewidth]{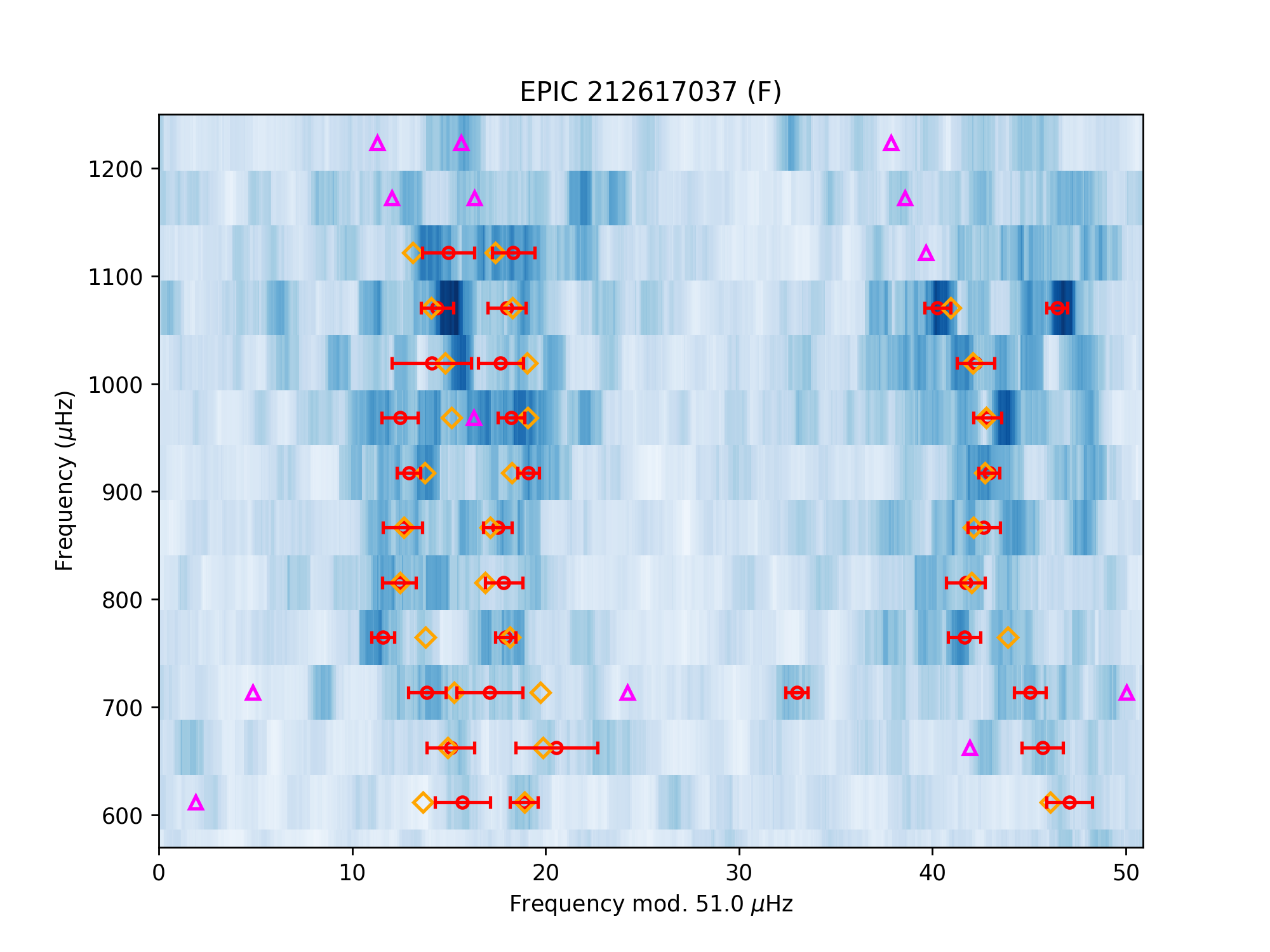}\\
    \includegraphics[width=0.5\linewidth]{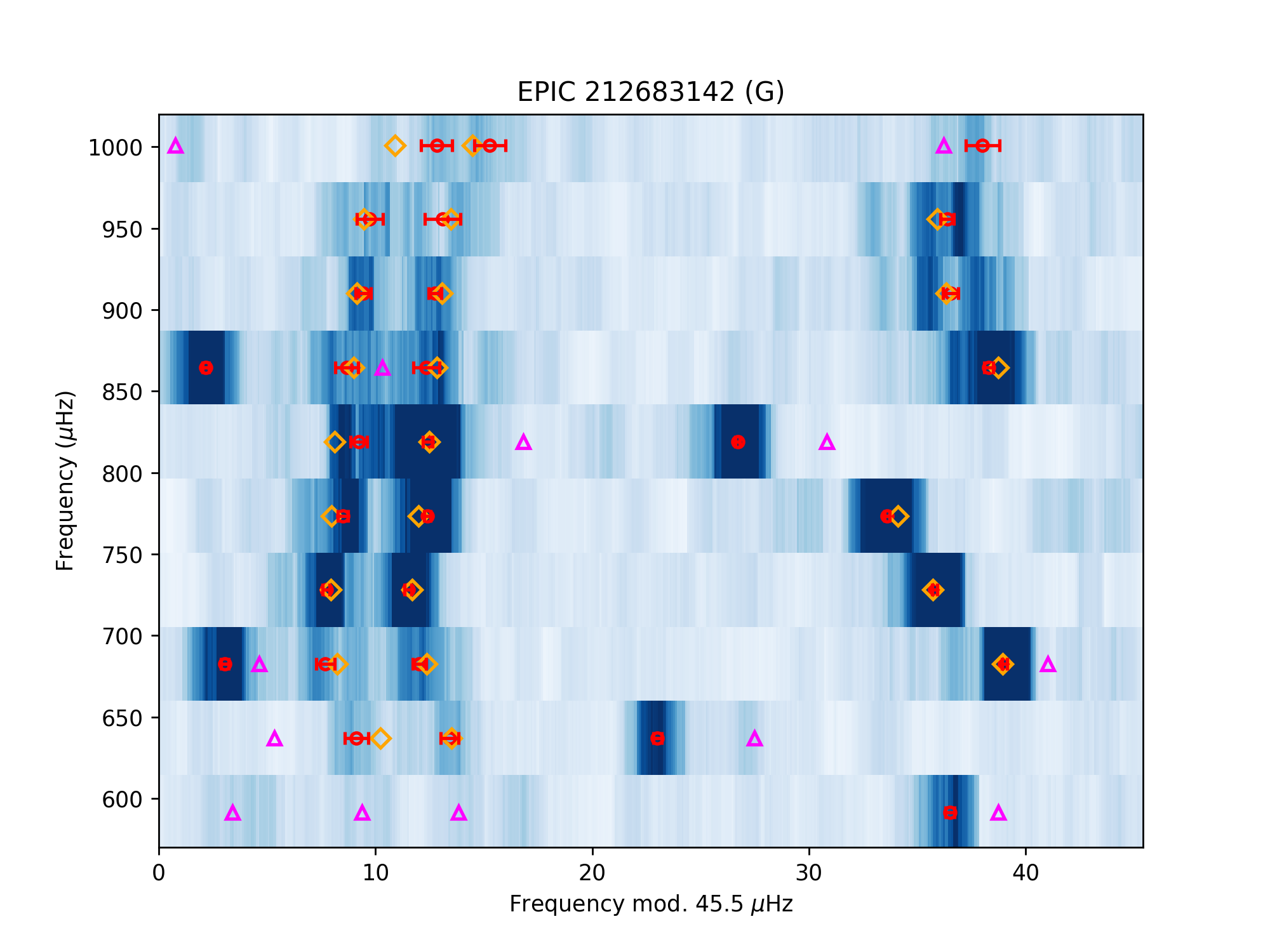}
    \includegraphics[width=0.5\linewidth]{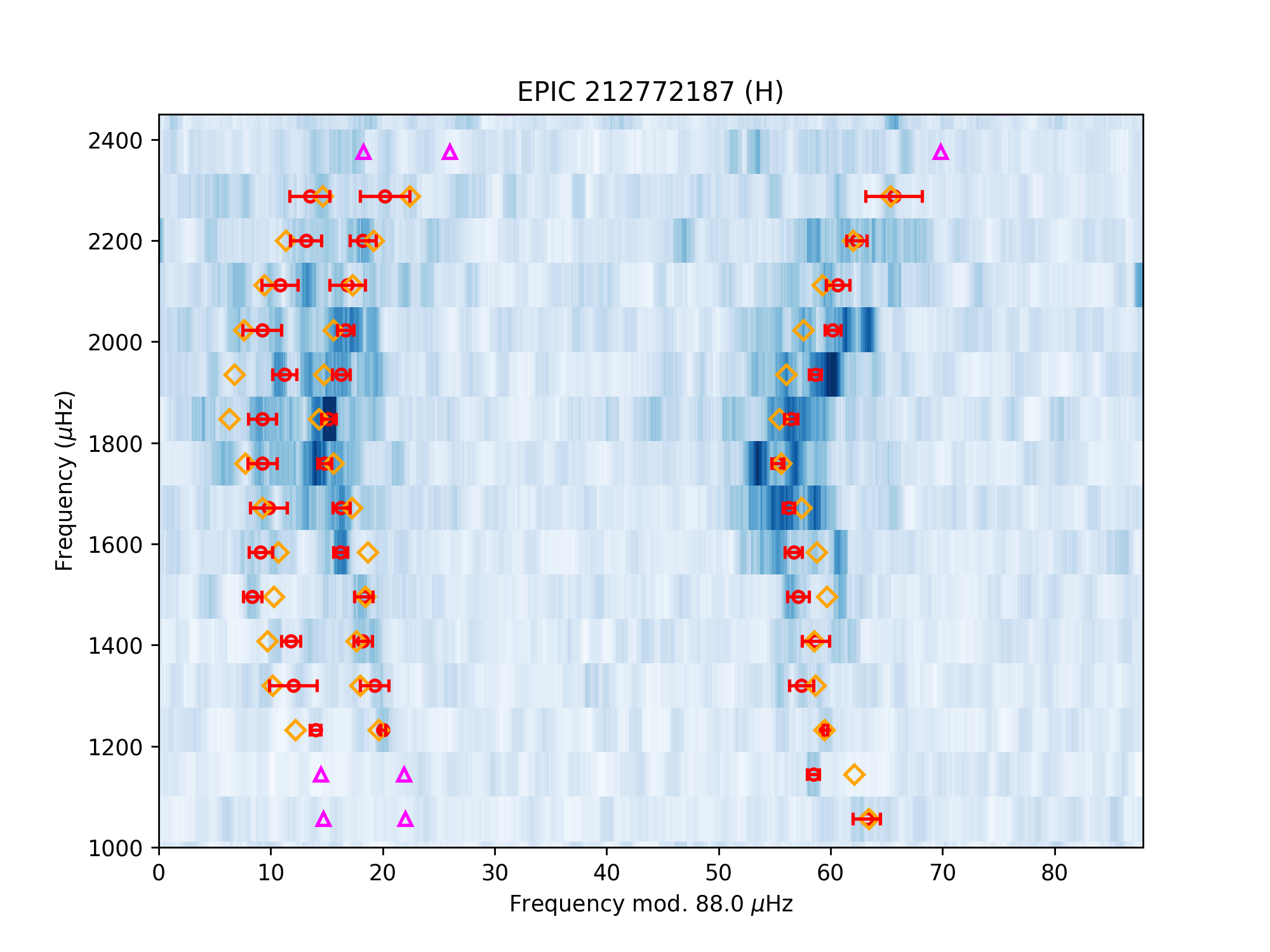}\\
    \caption{Echelle diagram of HD~120746 (E, EPIC~212516207) (top left panel), HD~117779 (F, EPIC~212617037) (top right panel), HD~119026 (G, EPIC~212683142) (bottom left panel) and HD~119038 (H, EPIC~212772187) (bottom right panel). Same legend as in Figure~\ref{fig:echelle_ABCD}.} 
    \label{fig:echelle_EFGH}
\end{figure}

\newpage

\section{Frequency comparison with \citet{2021ApJ...922...18O}}\label{appendix:comp_Ong}

\begin{figure*}[htb!]
    \includegraphics[width=0.48\linewidth]{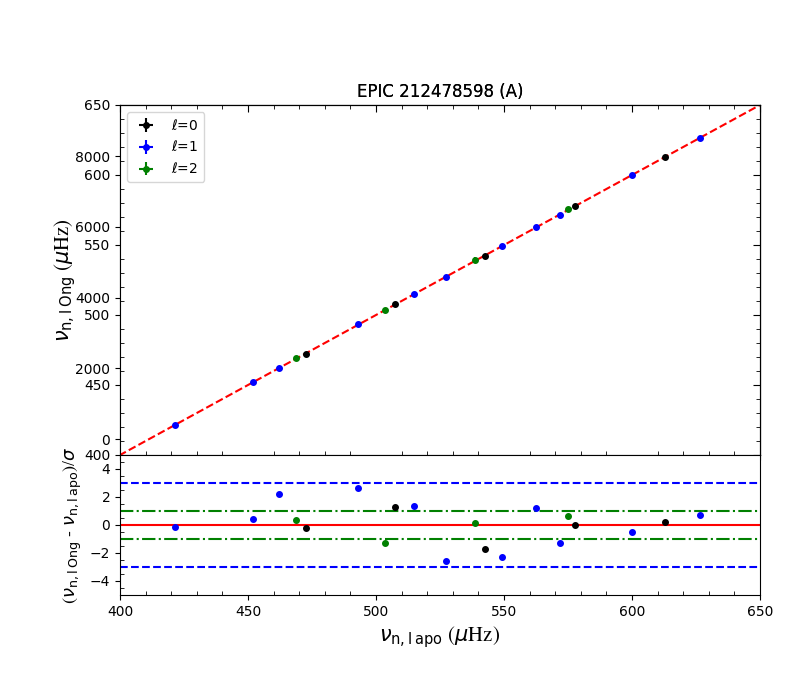}
    \includegraphics[width=0.48\linewidth]{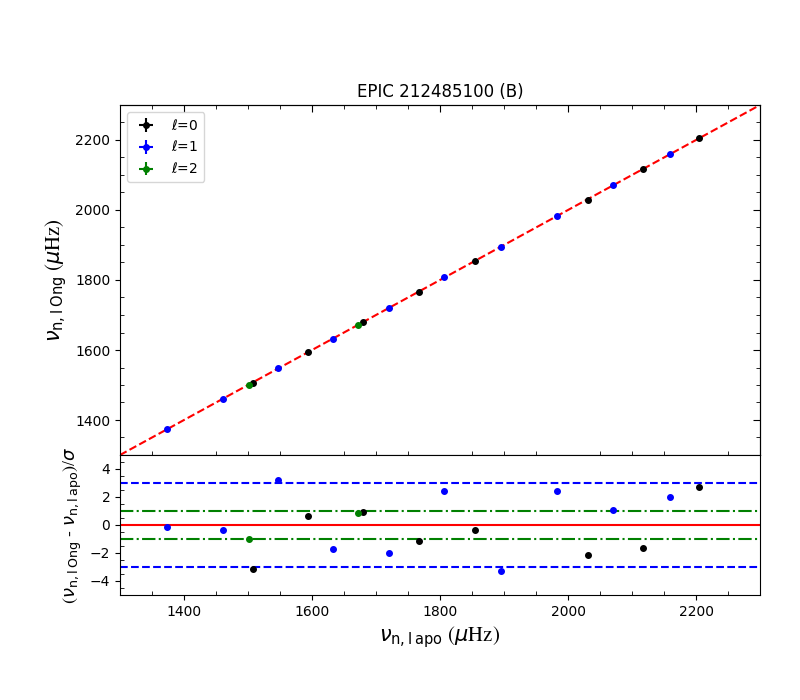}\\
    \includegraphics[width=0.48\linewidth]{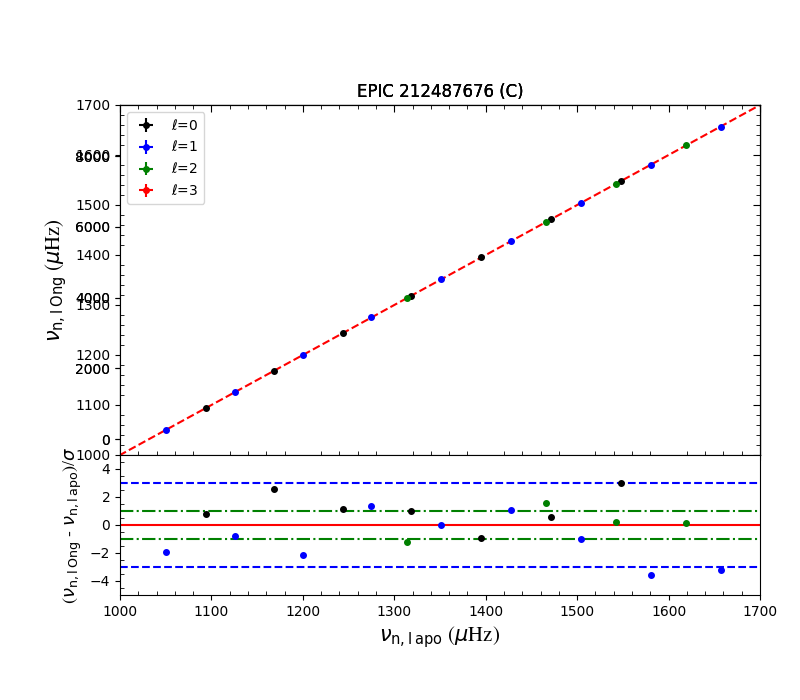}
    \includegraphics[width=0.48\linewidth]{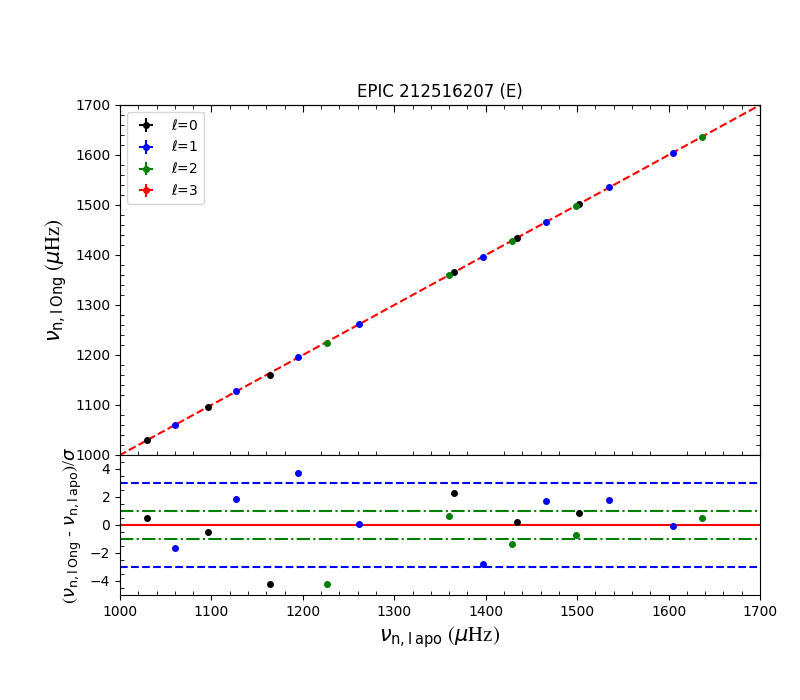}\\
    \includegraphics[width=0.48\linewidth]{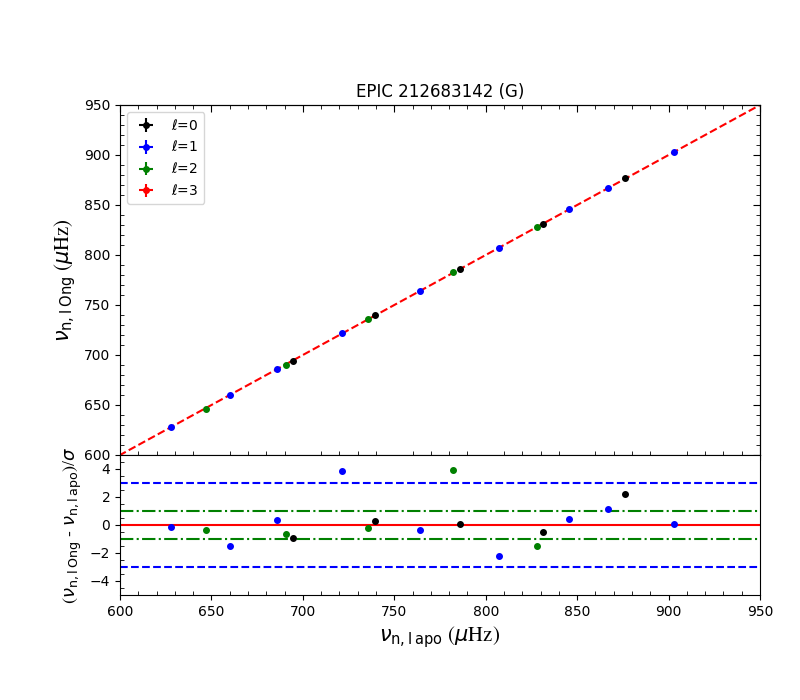}

    \caption{Comparison between the frequencies of modes in common obtained in this work ($\nu_{\rm n,l, apo}$) and those obtained in the analysis done by \citet{2021ApJ...922...18O} ($\nu_{\rm n,l, Ong}$). In the top panels, the red dashed line corresponds to the 1-to-1 line. In the bottom panels, we show the difference between the frequencies normalized by $\sigma$ that is computed as the square root of the quadratic sum of the frequency uncertainties from both methods. The green dashed-dotted (resp. blue dashed)  lines correspond to $\pm$\,1\,$\sigma$ (resp. $\pm$\,3\,$\sigma$).}
    \label{fig:comp_freq}
\end{figure*}

Since five of our targets (A, B, C, E, and G) were previously analyzed by \citet{2021ApJ...922...18O}, we compare their frequencies ($\nu_{\rm n,l, Ong}$) of the modes we have in common with the ones obtained with \texttt{apollinaire} ($\nu_{\rm n,l, apo}$) using two K2 campaigns. In Figure~\ref{fig:comp_freq}, we can see that there is a good agreement for the modes in common between both analyses. Most of the frequencies agree within 3\,$\sigma$ except a couple of modes for star E where the agreement is around 4\,$\sigma$.

\end{appendix}

\end{document}